\RequirePackage{nameref}
\documentclass{aa}  
\usepackage{keyval}

\usepackage{natbib}

\usepackage{verbatim}
\usepackage{stfloats}
\usepackage{xcolor}
\usepackage{adjustbox}
\usepackage{graphicx}
\usepackage{txfonts}
\usepackage{lipsum}
\usepackage{caption}

\DeclareSymbolFont{cmletters}{OML}{cmm}{m}{it}
\DeclareMathSymbol{v}{\mathalpha}{cmletters}{`v}
\usepackage{amsmath}
\usepackage{subcaption}          
\usepackage{lscape}
\usepackage{placeins}

\makeatletter
\long\def\@makecaption#1#2{
  \vspace{\abovecaptionskip}
  \sbox\@tempboxa{#1. #2}
  \ifdim \wd\@tempboxa >\hsize
    #1. #2\par
  \else
    \global \@minipagefalse
    \hb@xt@\hsize{\hfil\box\@tempboxa\hfil}
  \fi
  \vspace{\belowcaptionskip}}
\makeatother

\begin{document}

   \title{Dynamics of tidal dwarf galaxies in the system Arp 72}

    \author{Osvan M. Portilla-Narvaez\inst{1}
            \and
            Javier Zaragoza-Cardiel\inst{2,3}
            \and
            Gisela N. Ortiz-León\inst{1}
            \and
            Y. D. Mayya\inst{1}
            \and
            Beverly J. Smith\inst{4}
            \and
            Chandreyee Sengupta\inst{5,6}
            \and
            Mark L. Giroux\inst{4}
            \and
            S. Comerón\inst{7,8}
            \and
            Luz I. Alvarez-Cruz\inst{1}
            }

    \institute{Instituto Nacional de Astrofísica, Óptica y Electrónica,
              Luis Enrique Erro 1, Puebla, México\\
              \email{mauricio.portilla@inaoep.mx}
              \and
              Centro de Estudios de Física del Cosmos de Aragón (CEFCA), Plaza San Juan 1, 44001 Teruel, Spain
              \and
              Unidad Asociada CEFCA-IAA, CEFCA, Unidad Asociada al CSIC por el IAA y el IFCA, Plaza San Juan 1, 44001 Teruel, Spain
              \and
              East Tennessee State University, Department of Physics and Astronomy, Johnson City, TN 37614, USA
              \and
              Centre for Space Research, North-West University, Potchefstroom 2520, South Africa
              \and
              National Institute for Theoretical and Computational Sciences (NITheCS), South Africa
              \and
              Departamento de Astrofísica, Universidad de La Laguna, 38200 La Laguna, Tenerife, Spain
              \and
              Instituto de Astrofísica de Canarias, 38205 La Laguna, Tenerife, Spain
              }


   \date{Received September 30, 20XX}

 
  \abstract
   {Some interactions between galaxies produce tidal tails primarily composed of material from their disks. Within these tails, concentrations of gas and stars can form, resembling dwarf galaxies. These tidal objects often begin to form stars and become dynamically independent of their parent galaxies, leading to their classification as Tidal Dwarf Galaxies (TDGs). By definition, TDGs should consist solely of baryonic material, with a negligible dark matter fraction. In this study, we analyze the dynamics of two TDGs in the Arp\,72 system using high-resolution H$\alpha$ observations obtained with the MEGARA multi-spectrograph at the GTC (Gran Telescopio Canarias) and neutral hydrogen (H\,{\sc{i}}) data from the GMRT (Giant Metrewave Radio Telescope). The H\,{\sc{i}} data were also used to determine the gas mass. We derived the rotation curves and velocity dispersion from the kinematic data, which allowed us to estimate the dynamical mass of the systems. The TDGs were modeled through 3D fitting as rotating disks with additional pressure support, assuming a mass distribution following an exponential law. The gas mass was combined with the stellar mass to determine the total baryonic mass. Under specific dynamical considerations, we established the relationship between the dynamical mass and the baryonic mass. Additionally, we used the pressure-support corrected circular velocity to compare the behavior of TDGs in the context of the baryonic Tully-Fisher relation (BTFR) with the literature, showing how detached TDGs fall off the relation. The two studied objects are consistent with the expected properties of a TDG.}


   \maketitle
   \nolinenumbers

\section{Introduction}

    The interaction of galaxies can lead to the formation of tidal tails, where concentrations of gas and stars are likely to form. These structures can reach masses, sizes, and star formation rates comparable to those of dwarf galaxies \citep{1994A&A...289...83D, 10.1007/978-3-642-22018-0_37, Lelli2015, Gray_2023}. In contrast to dwarf galaxies of similar mass, tidal dwarf galaxies (TDGs) are expected to have a low dark matter content and a high metallicity \citep{1992Natur.360..715B, 2004A&A...427..803D, Bournaud2006, Lelli2015, 10.1093/mnras/stu2629, 2018MNRAS.474..580P, 2019A&A...626A..47H}. The low dark matter content arises because TDGs form from dynamically cold disk material, and after their formation, their gravitational potential is too weak to accrete significant amounts of dark matter from their surroundings. The high metallicity results from star formation occurring in pre-enriched material originating from the disks of the progenitor galaxies. Numerous candidate TDGs have been identified by their offset from the standard mass-metallicity relation for classical dwarf galaxies \citep{1992Natur.360..715B, 1994A&A...289...83D, 1998A&A...333..813D, 2004A&A...427..803D, Bournaud2006, 2009ApJ...705..723C, 2012A&A...538A..61M, 2014MmSAI..85..408S, Lelli2015, refId0_Javier2024}. \newline \indent
    TDGs represent a unique area of interest for studying galaxy formation, as they are currently the only galaxies in the process of forming. However, a key challenge for studying TDGs is their survival; their lifetimes must exceed 1 Gyr for their contribution to the total population of dwarf galaxies to be significant. \cite{1956ErNW...29..344Z} was the first to propose that collisions between galaxies could eject debris that would subsequently collapse gravitationally to form new small galaxies, implying that some dwarf galaxies may have a tidal origin. The lifetime of a TDG can be influenced by several factors, including tidal interactions with their parent galaxies, dynamical friction, and their own star formation activity. The survival rate of TDGs remains a topic of debate (see \citealp{2014MNRAS.440.1458D}). For instance, \citet{2000ApJ...543..149O} and \citet{10.1093/mnras/sts326} propose that almost all dwarf elliptical galaxies (dEs) could have a tidal origin, while \citet{10.1111/j.1365-2966.2012.21319.x} suggest that dwarf spheroidal galaxies (dSphs) and ultra-faint dwarf galaxies (UFDGs) near to the Milky Way might be ancient TDGs. On the other hand, \citet{2012MNRAS.419...70K} estimate that $\sim$6\% of dwarfs in clusters may have a tidal origin, \cite{10.1093/mnras/staa2985} estimate that 5\% of local dwarf galaxies are of tidal origin. \citet{Bournaud2006} and \citet{2019A&A...626A..47H} indicate that only a small percentage of TDGs evolve into independent dwarf galaxies. 
    
    In regular dwarf galaxies, the dark matter halo typically dominates even in the inner regions \citep{1988ApJ...332L..33C, 1997MNRAS.290..533D, 2011AJ....142...24O, 2015AJ....149..180O}. However, exceptions exist, such as dEs with stellar masses $M_{*}$\,>\,$10^{8}$\,M$_{\odot}$, which have little or no dark matter content in their inner regions \citep{10.1111/j.1365-2966.2011.18335.x}. In the outskirts of these dEs and in lower-mass dEs, the dynamical mass-to-light ($M/L$) ratios are often higher, though these measurements can be affected by insufficient suitable data and the dynamical assumptions employed \citep{10.1093/mnras/sts326}. The high $M/L$ ratios observed in dSphs and UFDGs, which make them seemingly incompatible with the nature of TDGs, might instead result from tidal effects \citep{10.1111/j.1365-2966.2012.21319.x}. A sample of ultra-diffuse galaxies (UDGs) with available H\,{\sc i} interferometric data reveals that they are predominantly baryon-dominated, indicating very low dark matter content \citep{ManceraPina2019}. These findings suggest a possible link between the origin of certain galaxy types and TDGs. Given the high frequency of galaxy interactions and mergers at high redshifts \citep{10.1093/mnras/279.3.L47, RyanJr2008, 2025MNRAS.539.1651F} and higher gas fractions at high redshifts \citep{2010ApJ...713..686D, 2010Natur.463..781T}, TDGs may have formed more frequently in the past \citep{2011ApJ...730....4B, 2018MNRAS.474..580P, 2019A&A...626A..47H}. Depending upon TDG survival rates, TDG formation may have been a significant alternate method for forming dwarf galaxies, in contrast to the standard hierarchical model in which baryonic matter settled into dark matter haloes in the early Universe, accreting additional gas over time via smooth flows and mergers.
    
    The formation of TDGs has been successfully reproduced in numerical simulations of galaxy mergers, which also account for gas dynamics and star formation \citep{10.1111/j.1745-3933.2008.00511.x}. These simulations reveal that in gas-rich galaxy mergers, Super Star Clusters (SSCs) and TDGs form simultaneously. High-resolution simulations allow the internal structure of TDGs to be resolved, showing that they are extended rotating disks with masses around $10^{8}$M$_{\odot}$. The internal dynamics of TDGs provide a way to determine the ratio between the dynamical and the baryonic masses of the system, thereby enabling an estimate of the non-baryonic dark matter content. This has been done for a limited sample of TDGs, as demonstrated by \citet{Braine2001}, \citet{Lelli2015}, \citet{roman2021}, and \citet{Gray_2023}. The kinematics of TDGs can be derived from various tracers, including the 21 cm line of neutral hydrogen (H\,{\sc i}), H$\alpha$ emission line, and CO lines. Line widths are frequently used to analyze their dynamics, as seen in studies like \citet{Gray_2023} and \cite{montes2024}. However, the recent formation and irregular nature of TDGs lead to diverse assumptions about their dynamics. For instance, they can be modeled as rotating disks with additional pressure support \citep{Lelli2015} or as spherically symmetric objects supported by velocity dispersion, as suggested by \citet{roman2021}. In some cases, kinematic distortions and warps can make determining rotation curves difficult \citep{2016MNRAS.455.2508S}. The variety of assumptions and methods introduces uncertainty in the determination of the ratio between the dynamical and the baryonic masses.\newline \indent
    The most suitable TDGs for dynamically resolved studies are detached TDGs, which are objects that are no longer connected to their parent galaxies by tidal structures \citep{refId0_Javier2024}. Since these objects tend toward a state of isolation, the kinematic tracers are less affected by the tidal field and surrounding material. In addition, measurements of the circular velocity provide important results on the location of TDGs in the baryonic Tully-Fisher relation (BTFR). The BTFR is an empirical relationship between the total baryonic mass ($M_{\rm{gas}}$$+$$M_{*}$) and the circular velocity in rotating galaxies. For gas-dominated galaxies, a robust determination of their position on the BTFR can be made, as the gas mass estimate is independent of the uncertainties associated with stellar mass-to-light ratios \citep{McGaugh_2012}. Objects with low dark matter content are located outside the BTFR. This result has been shown for TDGs by \cite{Lelli2015}, and for UDGs by \cite{ManceraPina2019}. In the latter case, it is unlikely that the UDGs in the sample originated from TDGs because they are isolated or very distant from any interacting system. Although there is considerable controversy about the origins of UDGs (see \citealp{2019MNRAS.484..245L, 2024OJAp....7E.117K, 2024A&A...687A.105I, 2025A&A...695A.124B}), there is less uncertainty about how dark-matter-free TDGs form. The existence of baryon-dominated TDGs are evidence for an alternative creation route for some dwarf galaxies.
    
    This article presents a study of two objects within the Arp\,72 system. Section \ref{Galaxy system and data} describes the system and the data used for the kinematic analysis. The methodology and kinematic model are detailed in Section \ref{Methods}. Results are presented in Section \ref{Results}, and discussed in Section \ref{Discussion}. Conclusions are provided in Section \ref{Conclusions}.

\section{Galaxy system and data}
\label{Galaxy system and data}
\subsection{Galaxy system}
\label{Galaxy system}

    Arp\,72 is a M\,51-like system composed of two interacting galaxies with a recession velocity close to 3295 km/s: NGC\,5996, the largest galaxy, and its companion NGC\,5994 are both barred galaxies classified as SBbc type \citep[HyperLEDA database][]{2014A&A...570A..13M}. The system exhibits a complex H\,{\sc i} distribution with tidal bridges, indicative of strong tidal interactions \citep{10.1111/j.1365-2966.2011.19767.x}. Additionally, it is considered a minor merger, with mass ratios estimates ranging from $\sim$\,0.1 to $\sim$\,0.5 \citep{10.1111/j.1365-2966.2011.19767.x}. The Arp\,72 system has the highest number of detected TDGs in the Spiral, Bridges and Tails (SB\&T) sample \citep{2016AJ....151...63S}, where four detached TDGs were characterized by \citet{refId0_Javier2024}, revealing ages greater than 300 Myr. Two of these TDGs exhibit stellar metallicities and nebular abundances higher than typically expected for dwarf galaxies. In the other two cases, the uncertainties are too large to definitively distinguish them from dwarf galaxies.

    For this work, we assume Arp\,72 to be a distance of 53.4\,$\pm$\,0.8\,Mpc (distance adopted for all calculations in this study), obtained assuming a Hubble-Lema\^{\i}tre constant $H_{0}$\,$=$\,73\,$\pm$\,1\,km/s/Mpc \citep{2019NatAs...3..891V} and a recession velocity 3902\,$\pm$\,27\,km/s (NASA/IPAC Extragalactic Database - NED, with the corrections by \citealp{2000ApJ...529..786M})

\begin{figure}
        \centering
        \includegraphics[width=\hsize]{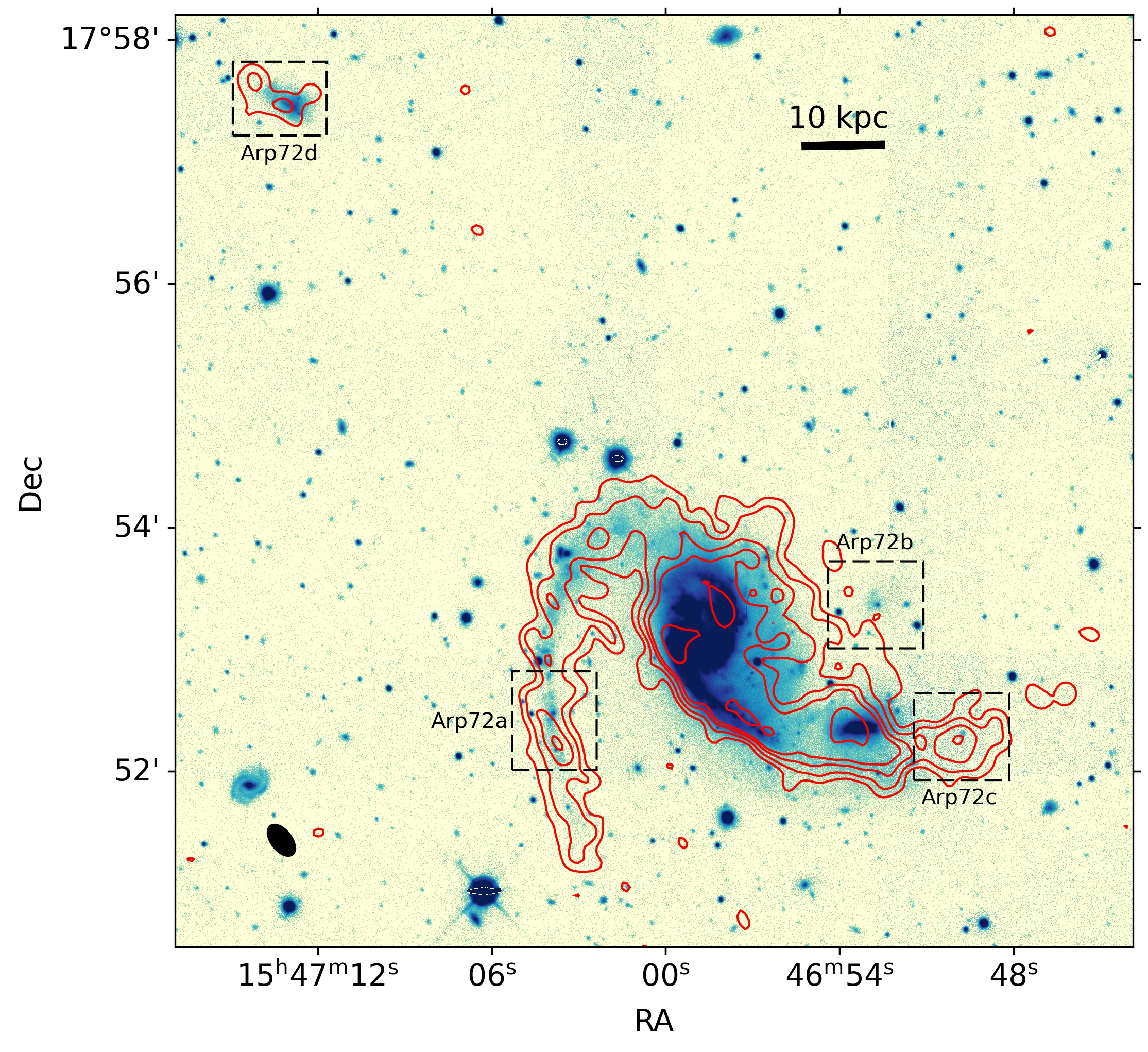}
        \caption{Location of the four detached TDGs identified by \cite{refId0_Javier2024} in the Arp\,72 system. The blue image corresponds to the DESI $g$-band, while the red contours represent the H\,{\sc i} column density map from GMRT data. Contours levels are 1.9, 4.6, 8.2, 11.0, and 21.3 $\times 10^{20}$ cm$^{-2}$.}
        \label{galaxy_desi_HI_contour}
    \end{figure}

    The four detached TDGs in Arp\,72, designed as Arp\,72a, Arp\,72b, Arp\,72c, and Arp\,72d, are marked in an optical image in Fig. \ref{galaxy_desi_HI_contour}. Arp\,72a and Arp\,72c are both located within the H\,{\sc i} tidal tails. Arp\,72b is situated close to the parent galaxies, while Arp\,72d is the most distant object in the system. Notably, neither Arp\,72b nor Arp\,72d show H\,{\sc i} tidal debris connecting them to Arp\,72, as illustrated in Fig. \ref{galaxy_desi_HI_contour}.


\subsection{Data}

    To study the optical morphology, we used a Hubble Space Telescope (HST) image in the $F606W$ filter and a $g$-band image from the Dark Energy Spectroscopic Instrument (DESI) \citep{2019AJ....157..168D}. We utilized GMRT (Giant Metrewave Radio Telescope) observations of the 21 cm neutral hydrogen line published in \cite{10.1111/j.1365-2966.2011.19767.x}, which were processed using {\sc aips} (Astronomical Image Processing System). Data affected by bad antennas, antennas with significantly lower gain, and radio frequency interference (RFI) were flagged. The data were then calibrated in amplitude and phase using primary and secondary calibrators. The primary flux density calibrator was 3C\,286, while the phase calibrator was J1609+266. The flux calibrator 3C286 was also used as the bandpass calibrator. The continuum was subtracted using the {\sc aips} tasks ``{\sc uvsub}'' and ``{\sc uvline}''. Finally, the {\sc aips} task ``{\sc imagr}''  was employed to generate a spectral cube. The characteristics of the data cube are summarized in Table \ref{properties_HI_cube}.
    \begin{table}[!ht]
    \caption{Properties of the H\,{\sc i} datacube.}
    \centering
    \begin{tabular}{ccccc}
    \hline
    \begin{tabular}[c]{@{}c@{}} $\Delta v$ \\ {[}km/s{]} \end{tabular} & \begin{tabular}[c]{@{}c@{}} $B_{\rm maj}$ \\ {[}arcsec{]} \end{tabular} & \begin{tabular}[c]{@{}c@{}} $B_{\rm min}$ \\ {[}arcsec{]} \end{tabular} & \begin{tabular}[c]{@{}c@{}} rms noise \\ {[}mJy/beam{]} \end{tabular} & \begin{tabular}[c]{@{}c@{}} PA \\ {[}$^{\circ}${]} \end{tabular} \\ \hline \hline
     13.5 & 18 & 10 & 0.6 & 37.8
    \\ \hline
    \end{tabular}
    \tablefoot{$\Delta {v}$ is the channel width, $B_{\rm maj}$ and $B_{\rm min}$ are the half-power beam widths along the major and minor axes, respectively, and PA is the beam position angle.}
    \label{properties_HI_cube}
    \end{table}
    
    Additionally, for two TDGs (Arp\,72b and Arp\,72c), we obtained H$\alpha$ data from MEGARA (Multi Espectrógrafo en GTC de Alta Resolución para Astronomía) \citep{10.1117/12.2313040} at the Gran Telescopio Canarias (GTC) in the HRR band in the IFU mode, whose field of view is 12.5'' $\times$ 11''. The spectral resolution was $\Delta v_{\rm{FWHM}} = 12.9$ km/s (measured from the calibration lamps) with a sampling of 0.097\,\AA\,(4.4 km/s). MEGARA is equipped with 623 hexagonal fibers of 0.62 arcsec diameter with the spatial resolution determined by the seeing which was 1.1 arcsec. These observations were carried out on the night of 2024-05-06, 07 under spectroscopic sky conditions as part of the GTC proposal GTCMULTIPLE4B-24A - 4440702, 4441018 (PI Javier Zaragoza-Cardiel), and were processed using the standard MEGARA pipeline \citep{sergio_pascual_2018_2206856}, with the two wavelength calibration lamps combined for improved accuracy. To reduce the spectrum noise and smooth the line profile in the MEGARA spectra, we use binomial smoothing:
    \begin{equation}
        F_{i}=0.25F_{i-1}+0.5F_{i}+0.25F_{i+1}
    \end{equation}
    \noindent where $F_{i}$ is the flux in the $i$-th pixel. This was done in order to improve the results of the fits described in section \ref{Kinematic model and 3D fit} without losing much spectral resolution, which is crucial for this work.

\subsection{TDGs}
    
    Arp\,72b is a system where H\,{\sc i} is not detected. In contrast, Arp\,72a and Arp\,72c are prominent features within the tidal arms, and are seen as gas overdensities. Arp\,72a, Arp\,72b, and Arp\,72c show active star formation evidenced by their significant brightness in UV images \citep{Smith_2010, 10.1111/j.1365-2966.2011.19767.x}, and confirmed by their optical spectra \citep{refId0_Javier2024}. The H\,{\sc i} emission in Arp\,72b might be below the detection limit of the GMRT. As for Arp\,72d, it is located approximately 5.6 arcmin from the center of NGC\,5996. Using the distance from Section \ref{Galaxy system}, Arp\,72d is at a projected distance of around 87 kpc from the center of the system. The significant separation and the absence of tidal debris connecting Arp\,72d to either NGC\,5996 or NGC\,5994 challenge its classification as a tidal object. The stellar masses and ages of the TDGs in Arp\,72, as estimated by \citet{refId0_Javier2024}, are presented in Table \ref{stellarmass_age_table}.

    The analysis presented hereafter is restricted to Arp\,72b with H$\alpha$ and Arp\,72c with H\,{\sc i}. The H$\alpha$ data from MEGARA for Arp\,72c do not have enough fibers detected with a good signal-to-noise ratio to allow a reliable 3D fit. For Arp\,72a and Arp\,72d, data with a higher signal-to-noise ratio and improved spatial resolution are required.

    \begin{table}[!ht]
    \caption{Properties of the TDGs sample.}
    \centering
    \begin{adjustbox}{width=\linewidth, center}
    \begin{tabular}{cccccc}
    \hline
    Object & \begin{tabular}[c]{@{}c@{}}$M_{*}$\\ {[}$10^{7}$ M$_{\odot}${]}\end{tabular} & \begin{tabular}[c]{@{}c@{}}Age\\ {[}Myr{]}\end{tabular} & RA & Dec & Ref. \\ \hline \hline
    Arp\,72a & $1.2 \pm 0.4$ & $430 \pm 150$ &  15:47:03.9 & +17:52:28.7 & a \\
    Arp\,72b & $1.7 \pm 0.5$ & $830 \pm 330$ & 15:46:52.7 & +17:53:22.1 & a \\
    Arp\,72c & $0.25 \pm 0.13$ & $320 \pm 110$ & 15:46:49.8 & +17:52:19.1 & a \\
    Arp\,72d & $8.4 \pm 2.7$ & $870 \pm 330$ & 15:47:12.8 & +17:57:26.8 & a \\ \hline
    \end{tabular}
    \end{adjustbox}
    \tablefoot{Stellar mass and ages of the TDGs. The age corresponds to the time since the last star formation burst. We report the coordinates of the peak optical emission in the DESI $g$-band image.}
    \tablebib{(a) \citet{refId0_Javier2024}.}
    \label{stellarmass_age_table}
    \end{table}

\section{Methods}
\label{Methods}
\subsection{Dynamics of Tidal Dwarf Galaxies}
\label{Dynamics of Tidal Dwarf Galaxies}

    TDGs are gas-dominated systems, typically exhibiting gas-to-stellar mass ratios greater than two. The majority of TDGs show $M_{\rm{H}\,\text{\sc i}} /M_{*}$\,$\sim$\,10 \citep{Lelli2015, roman2021, Gray_2023}. This indicates that gas is the dynamically dominant component in these systems. Building upon the results from \cite{10.1111/j.1745-3933.2008.00511.x} and \cite{Lelli2015}, we model TDG dynamics using H\,{\sc i} and H$\alpha$ observations and assuming pressure-supported rotating disks with mass distributions following an exponential law.

    The dynamical model is detailed in Appendix \ref{Dynamical model append}. In our methodology, the dynamics of the objects are defined by the disk rotation, $v_{\rm rot}(R)$, where $R$ is the galactocentric radius in the disk plane. The rotation is expressed as a function of the central surface mass density, $\Sigma_{0}$, and the disk scale length, $R_{\rm d}$. The circular velocity, $v_{\rm c}$, at a given radius depends on $v_{\rm rot}$ and the velocity dispersion, $\sigma_{v}$. Regarding the structure of the objects, for the H\,{\sc i} component, the surface brightness distribution $\mu_{\rm{H}\,\text{\sc{i}}}$ depends on $R$, $R_{\rm d}$, and $R_{\rm c}$, where the latter represents the cut radius of a truncated profile (Eq. \ref{HI_profile_ecuacion}). For the MEGARA data, the $\mu_{\rm{H}\alpha}$ surface brightness distribution was derived from the optical images. The fitting procedure for these quantities, along with other parameters, is described in Sect. \ref{Kinematic model and 3D fit}. Our 3D fits incorporated data masking and weighting as detailed in Sect. \ref{Masking and weighting of data}, while the uncertainties were calculated following the procedure outlined in Appendix \ref{Determination of uncertainties}.

\subsection{Kinematic model and 3D fit}
\label{Kinematic model and 3D fit}

    The kinematic fitting procedure starts with {the} 2D modeling of the velocity field through $\chi^{2}$ minimization. We divided the velocity field into concentric elliptical annuli, with their geometry defined by the kinematic center ($x_{\rm{kc}}$, $y_{\rm{kc}}$), inclination ($i$), and position angle ($\varphi_{0}$) of the galaxy. During the search of the optimal values the above parameters are varied in the outermost loops. In each iteration, an appropriate number of rings is defined in which we first simultaneously fit both the systemic velocity ($v_{\rm{sys}}$) and rotational velocity ($v_{\rm{rot}}$). We then performed a second step where we refined the rotational velocity profile while fixing $v_{\rm{sys}}$ to the mean value derived from all rings. The equation fitted to each ring is:
    \begin{equation}
        {v}_{\rm{obs}} = {v}_{\rm{sys}} + {v}_{\rm{rot}} \sin{i} \cos{\theta} \, .
        \label{vrot2D}
    \end{equation}
    The coordinate transformation implied by Eq. \ref{vrot2D} is described by \cite{1978ARA&A..16..103V}. This 2D fit gives a value for the parameters $x_{\rm{kc}}$, $y_{\rm{kc}}$, $\varphi_{0}$, $i$, $v_{\rm{sys}}$, and $v_{\rm{rot}}$ which are then used later to give initial values to the parameters in the 3D fit.
    
    Just as the rotation curve $v_{\rm{rot}}(R)$ is affected by instrumental resolution effects (beam for radio observations or PSF for optical data), the measured velocity dispersion in each spaxel (or fiber) includes \cite[see their appendix B]{10.1093/mnras/stx3016}:
    \begin{itemize}
        \item spectral broadening due to beam-smearing effects, and 
        \item in optical observations, additional convolution with the instrumental line spread function (LSF).
    \end{itemize}
    A methodology to address this problem is 3D spectral cube fitting, in which a spectral cube obtained from a disk model and convolved with instrumental effects is incorporated during parameter optimization. This process provides a self-consistent treatment of resolution effects \citep[e.g.][]{10.1093/mnras/stv1213}. Another alternative is to correct the measured quantities by characterizing the instrumental effects with models convolved with the telescope response \citep[e.g.][]{10.1093/mnras/stx3016}. We developed a {\sc python} code to generate a galactic disk model based on the brightness distributions described in Section \ref{Dynamics of Tidal Dwarf Galaxies}, assuming uniform velocity dispersion throughout the disk. The disk rotation in the ideal case is derived from Eq.\ref{vrot} which is then projected using Eq. \ref{vrot2D}, thus the position of the Gaussian centroid is determined on the spectral axis at each spatial location. Two methods were used:
    \begin{enumerate}
        \item Parametric rotation curve: the rotation of the disk is obtained from the fit of the parameters $\Sigma_{0}$ and $R_{\rm{d}}$ (Eq. \ref{vrot}). That is, the observed velocity map is linked to these two parameters.
    \end{enumerate}
    Alternatively, when it was not possible to fit the values of $\Sigma_{0}$ and $R_{\rm{d}}$, the rotation was fitted as follows:
    \begin{enumerate}
    \setcounter{enumi}{1}
        \item Fixed rings: the rotation of the galaxy is obtained by fitting a value of  $v_{\rm{rot}}$ to each ring using Eq. \ref{vrot2D}, which produces the rotation curve that is then used to derive $\Sigma_{0}$ and $R_{\rm{d}}$ through a fitting of the Eq. \ref{vrot}.
    \end{enumerate}
    In both cases, the projection onto the image plane was performed using Eq. \ref{vrot2D}, where the observable is $v_{\rm{obs}}$ (moment 1, defined in Eq. \ref{ecuaciones_momentos}, i.e., the centroid of the line). This projection allows the model spectra to be compared with observations and thus fit the aforementioned parameters. We choose between these fitting approaches based on the quality of the fit to all parameters, where in the optimal case, the rotation curve can be constrained by $\Sigma_{0}$ and $R_{\rm{d}}$.
    
    This procedure yields a model spectral cube that is subsequently convolved with the instrumental parameters. However, the H\,{\sc i} and H$\alpha$ data have distinct architectures. For the H\,{\sc i} data, we generated a spectral cube, whereas for H$\alpha$ we extracted individual fiber spectra. The main parameters to be fitted are then $\Sigma_{0}$, $R_{\rm{d}}$, $x_{\rm{kc}}$, $y_{\rm{kc}}$, $\varphi_{0}$, $i$, $v_{\rm{sys}}$, and $\sigma_{v}$ which are set constant throughout the disc. Thus, a 3D-fit was performed to obtain the parameters of the disk with $\chi^{2}$ minimization with the {\sc lmfit} package\footnote{\url{https://lmfit.github.io/lmfit-py/}} \citep{newville_2025_16175987}. Before minimizing $\chi^{2}$, the line profiles in each spaxel or fiber are normalized because we are mainly interested in the value of $\sigma_{v}$ from the line profile. We emphasize that the normalization of the spectra does not remove the relevance of the H\,{\sc i} and H$\alpha$ profiles during the fitting, because the line intensity distribution is taken into account by the model spectral cube at the time of convolution. For the analysis presented in Sect. \ref{Results}, the kinematic center was manually determined through visual inspection of the velocity map, thereby reducing by two the number of free parameters in the minimization.

\subsection{Morphology, observed kinematics, and baryonic mass}

    The structure of TDGs exhibits irregular features characteristic of dwarf galaxies, where the distributions of ionized and neutral gas do not necessarily coincide \citep{Lelli2015}. Consequently, kinematic tracers likely provide more reliable estimates of the total mass distribution, with consistency expected between the kinematic profiles derived from different tracers \citep{10.1093/mnras/stad2790}. In our study, we can assess kinematic consistency only for Arp\,72c, where we have H$\alpha$ data and detection of its counterpart in H\,{\sc i}.
    
    We constructed the intensity ($M_{0}$), velocity ($M_{1}$), and velocity dispersion ($M_{2}$) maps by computing the spectral moments of the emission lines, using only channels where the line flux exceeded the noise level. The calculation was performed using:
    \begin{equation}
        \begin{split}
            & M_{0} = I_{\rm{obs}} = \int I_{v} dv \, , \\
            & M_{1} = {v}_{{\rm obs}} = \dfrac{\int {v} I_{{v}} dv}{M_{0}} \, , \\
            & M_{2} = \sigma_{v}^{2} = \dfrac{\int ({v}-M_{1})^{2} I_{v} d{v}}{M_{0}} \, , 
        \end{split}
        \label{ecuaciones_momentos}
    \end{equation}
    \noindent here, $I_{v}$ is the intensity in the channel with velocity $v$ and $dv$ is the channel width for H\,{\sc i} observations and spectral sampling for H$\alpha$ observations.
    
    These equations are independent of Gaussian profile assumptions. Consequently, the 3D fitting results may show significant differences with the data in the velocity maps. This occurs because $v_{\rm{obs}}$ represents the intensity-weighted velocity, whereas the 3D fit assumes a Gaussian profile centered on the spectral axis according to the disk rotation model, which will then be affected by the instrumental parameters.
    
    Similarly to \cite{Haynes_2018}, the H\,{\sc i} mass was calculated from
    \begin{equation}
        M_{{\rm{H}\,\text{\sc{i}}}}[{\rm M}_{\odot}] = 2.356\times 10^{5} \left [\frac{D^{2}}{{\rm Mpc}^{2}} \right] \left [ \frac{F_{\rm{H}\,\text{\sc{i}}}}{\rm{Jy} \cdot \rm{km/s}} \right] \, ,
        \label{HI_mass}
    \end{equation}
    \noindent where $D$ is the distance to the object in Mpc and $F_{\rm{H}\,\text{\sc{i}}}$ is the flux of the 21 cm line. To take into account helium and heavier elements, the mass of neutral gas must be multiplied by a factor between 1.33 and 1.4 depending on the nature of the object \citep{McGaugh_2012}. For this work, we use $M_{\rm{gas}}$=$1.4M_{\rm{H}\,\text{\sc{i}}}$. Then, the baryonic mass is $M_{\rm{bary}}$=$M_{\rm{gas}}$+$M_{*}$. We assume a small or zero contribution of molecular gas mass due to the lack of CO observations, and take this assumption into account when presenting the results.

\section{Results}
\label{Results}

\subsection{Delimitation of TDGs, masking and weighting of data}
\label{Masking and weighting of data}

\begin{figure*}[!ht]
        \begin{subfigure}[b]{0.55\textwidth}
            \includegraphics[width=\textwidth]{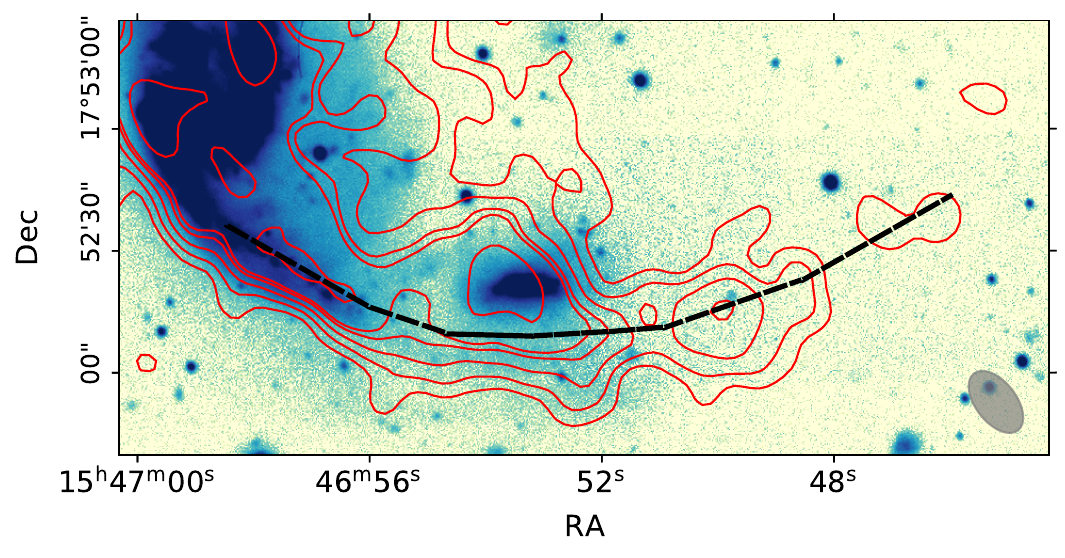}
            \label{tidal:img1}
        \end{subfigure}
        \hfill %
        \begin{subfigure}[b]{0.4\textwidth}
            \includegraphics[width=\textwidth]{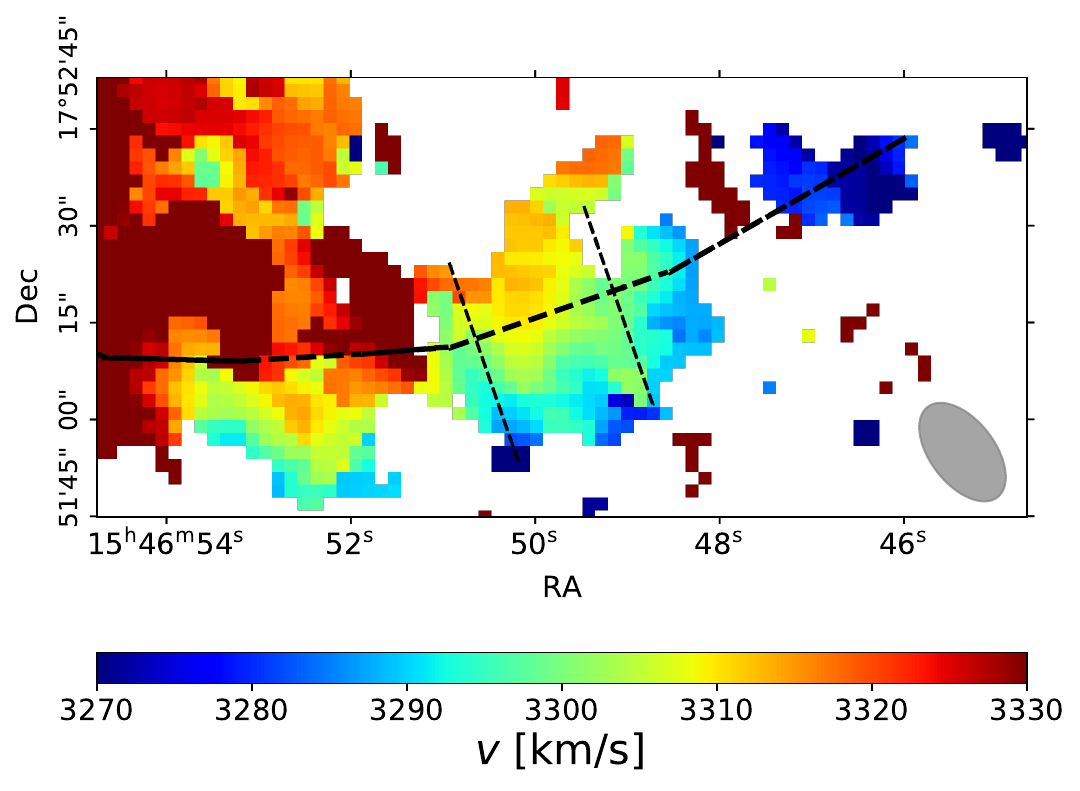}
            \label{tidal:img2}
        \end{subfigure}
        \vspace{0.01cm}
    \begin{minipage}[c]{0.625\textwidth}
        \begin{subfigure}[c]{\textwidth}
            \includegraphics[width=\textwidth]{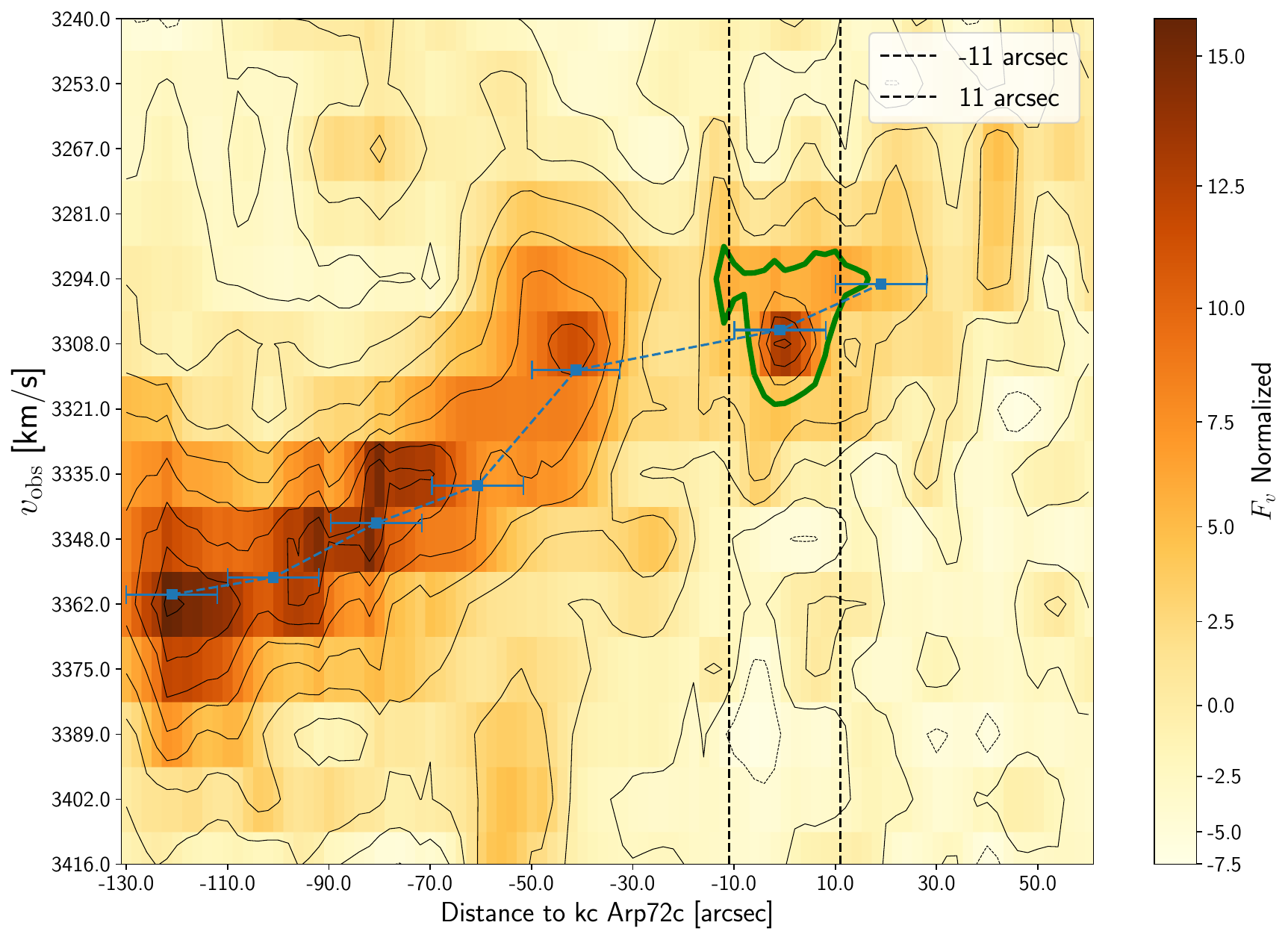}
            \label{fig:img3}
        \end{subfigure}
    \end{minipage}
    \hfill
    \begin{minipage}[c]{0.325\textwidth}
        \caption{Trajectory and PV diagram of the tidal tail in Arp\,72. Top panels: The trajectory (black dashed line) is shown over the DESI $g$-band image with red H\,{\sc i} contours (left) and the H\,{\sc i} velocity map (right). Perpendicular lines indicate the TDG boundaries derived from the PV analysis. Bottom panel: PV diagram covering the entire tidal structure. Blue squares represent the tidal debris gradient calculated via flux-weighted binning. The spatial offset is relative to the kinematic center (kc) of Arp\,72c. The green contour highlights the TDG boundary (SNR $\sim 5$), and the black dashed vertical lines mark its spatial extent.}
        \label{tidal:3_images}
    \end{minipage}
    \end{figure*}

    Arp\,72c belongs to the south-eastern tidal tail of Arp\,72; therefore, it is necessary to isolate the material associated with the TDG from the surrounding tidal debris. To this end, we constructed the Position-Velocity (PV) diagram shown in Fig. \ref{tidal:3_images}. The top-left panel shows the trajectory used to build the PV diagram overlaid on the DESI $g$-band optical image with H\,{\sc i} contours. The top-right panel shows the H\,{\sc i} velocity map of the tidal tail and the delimitation of the TDG along the direction of the tidal arm. The bottom panel presents the PV diagram, where blue squares indicate the gradient traced by the highest gas density regions. The green contour represents the boundary used to define the TDG, with the 11-arcsecond limits (see below) approximately determined by this contour.

    The PV diagram shows that the velocity gradient of the tidal arm decouples starting at $-30$ arcsec, with a clear change in slope, suggesting that Arp\,72c tends to be isolated from any larger kinematic structure. The contour selected to delimit the TDG has a signal-to-noise ratio of $\sim 4.7$. Given the asymmetries at $v_{\rm obs}=3294$ km/s, we reduced the horizontal extent to 11 arcseconds (indicated by dashed vertical lines). This limit is also shown on the velocity map, where it can be seen that the velocity field becomes more irregular beyond this boundary, which may provide evidence of material that is not gravitationally bound to the TDG.

    The case of Arp\,72b is easier as it is not associated with any tidal structure; therefore, the analysis was performed across the entire region where H$\alpha$ emission was detected.
    
    The observational data were fitted only at positions with line detections exceeding a specified signal-to-noise threshold. For the H\,{\sc i} data, additional masking was applied based on a visual inspection of the velocity field to exclude tidal debris near the optical positions of the TDGs. Along the spectral axis, we assigned reduced weights to channels below the noise level in order to focus solely on the detected emission. The high spectral resolution of the H$\alpha$ data allows us to detect double-peak profiles, so in this case we applied weighting to give priority to the main peak, which is equivalent to weighting by intensity. On the other hand, for H\,{\sc i} data, no weighting by intensity was performed on the channels with signal.

\subsection{Surface brightness and mass profiles}

\begin{figure}[!b]
        \centering
        \includegraphics[width=\hsize]{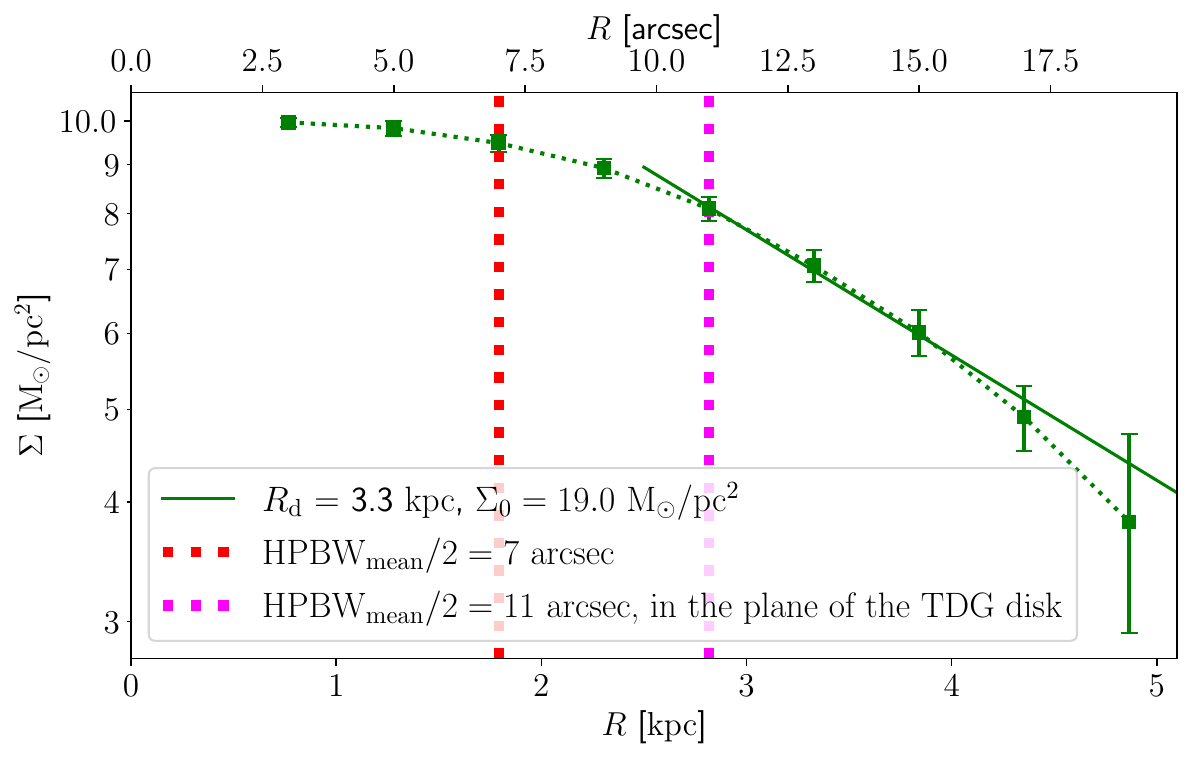}
        \caption{Mass surface density profile of Arp\,72c obtained from the H\,{\sc i} intensity map using the geometric parameters of the 3D fit (squares connected by dotted line). The fit of an exponential profile to the outermost radii is also shown (solid line). The vertical red line indicates half of the mean beam size (the average of $B_{\rm min}$ and $B_{\rm maj}$) and the magenta line corresponds to 7 arcseconds in the plane of the TDG disk, marking the threshold beyond which spatially independent data points exist.}
        \label{HI_profiles}
    \end{figure}

\begin{figure}[!t]
        \centering
        \includegraphics[width=\hsize]{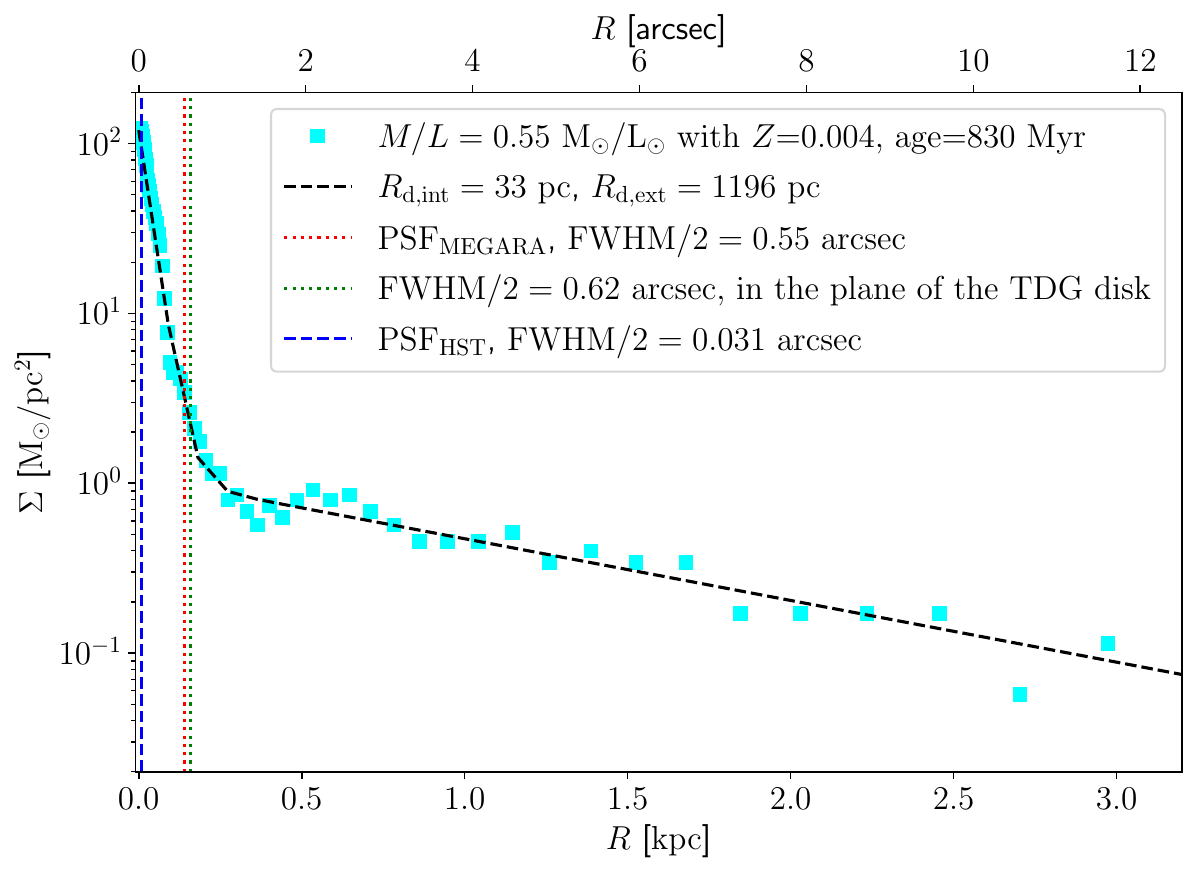}
        \caption{Optical profile of Arp\,72b obtained with the HST $F606W$ filter, with their respective combined exponential fit in the dashed black line. The profile was transformed into mass surface density units using the mass-to-light ratio from models of \cite{Bruzual2003} with an initial mass function (IMF) of Kroupa. The blue vertical dashed line is half of the PSF FWHM for HST. Red and green (in the plane of the TDG disk) vertical dotted lines denote half of the FWHM for MEGARA-GTC, marking the radius beyond which spatially independent data points exist.}
        \label{F606W_profiles}
    \end{figure}

    For Arp\,72c using the $\varphi_{0}$, $i$, and ($x_{\rm{kc}}$, $y_{\rm{kc}}$) values derived from the 3D fit results, we obtained the H\,{\sc i} profile from intensity maps ($M_{0}$ in Eq. \ref{ecuaciones_momentos}) with the {\sc iraf-ellipse} task. This profile was expressed as H\,{\sc i} mass surface density. We additionally obtained the surface brightness profile of Arp\,72b using the HST $F606W$ filter, with the profile center fixed at the brightest pixel. These measurements enabled direct comparison between the kinematic results and observed brightness distributions, primarily allowing us to evaluate the derived $R_{\rm{d}}$ values and estimate their associated uncertainties. The mass surface density profile derived from the H\,{\sc i} distribution is presented in Fig.\,\ref{HI_profiles} for Arp\,72c, while the optical surface brightness profile transformed into mass surface density profile of Arp\,72b is shown in Fig. \ref{F606W_profiles} (using the models of \citealp{Bruzual2003} with metallicities and ages closest to those obtained by \citealp{refId0_Javier2024}). The nominal spatial resolution of the various observations FWHM$_{\rm MEGARA}=1.1$\,arcsec and HPBW$_{\rm GMRT,\,mean}=14$\,arcsec, might suggest a limited spatial sampling. However, the profiles extracted via IRAF and our 3D fitting procedure utilize points separated by at least one resolution elements on opposite sides of the disk at a given radius $R$. Consequently, for radii greater than half the spatial resolution, the determination of mass and kinematic profiles is based on independent data points, from which parameters can be derived. To ensure a consistent comparison across all positions, particularly along the minor axis, we also show the spatial resolution projected onto the disk plane (vertical dotted lines marked in Figs. \ref{HI_profiles}, \ref{F606W_profiles}, \ref{rotation_dispersion_arp72b_new_version}, \ref{rotation_dispersion_arp72c_new_version} and \ref{all_profiles_mass.}). In the 3D fitting process, resolution effects are mitigated by convolving the model with the synthesized beam or PSF prior to comparison with the observations. While this approach facilitates the extraction of parameters from the available datasets, the correlation between adjacent points remains present.
    
    As specified in Eq. \ref{HI_profile_ecuacion}, we fitted an exponential profile to the H\,{\sc i} mass surface density profile (Fig. \ref{HI_profiles}) only in regions showing an exponential fall. The latter gives an estimation of $\Sigma_{0}$ and $R_{\rm d}$.

    For Arp\,72b, where optical spectral data enabled kinematic analysis, we fitted the brightness profiles in Fig. \ref{F606W_profiles} with dual exponential components. The HST $F606W$ filter resolves at least two distinct structures: (i) a compact component with scale lengths $R_{\rm{d,int}}$\,<\,100\,pc, likely corresponding to star-forming regions, and (ii) an extended component characterized by $R_{\rm{d, ext}}$. The latter may include both diffuse ionized gas and optical emission from tidal material, given the spectral response of the filter. The central region of Arp\,72b distinguishes itself with a surface mass density approximately $\sim$10 times higher than Arp\,72c. Furthermore, given the value of $R_{\rm d,\, int}$, it is more compact than Arp\,72c.

\subsection{Kinematics 3D fits}
\label{Kinematics fits}

    We applied the methodology described in Sect. \ref{Kinematic model and 3D fit}, following the considerations detailed in Sect. \ref{Masking and weighting of data}, while acknowledging that our results may be limited by the observational data quality. Both the signal-to-noise ratio (SNR) and spatial resolution impose significant limitations; specifically, at positions furthest from the kinematic center, the SNR tends to drop to values around 3. Furthermore, TDGs are intrinsically compact objects. In the case of Arp\,72c, for instance, our delimitation via PV diagrams and velocity map inspection resulted in only two resolution elements along each kinematic axis; consequently, Arp\,72c is marginally resolved.

    Discrepancies between the observed and modeled velocity dispersion maps are expected, as our model assumes an intrinsically uniform $\sigma_{v}$ throughout the disk. Within the model, any spatial variations in the $\sigma_{v}$ field arise exclusively from beam-smearing effects; consequently, the best-fit $\sigma_{v}$ represents a characteristic average across the entire disk.

    Another apparent limitation is the restricted number of spatial resolution elements within the sources, which is particularly evident in the H\,{\sc i} modeling of Arp\,72c. In this case, there are at most four independent spatial points (approximately two beams along both the minor and major axes given the beam orientation), which would nominally preclude fitting more than three parameters. However, since we perform a 3D fit, the emission line at each spatial position spans at least three spectral channels above the noise level in nearly all spaxels. A similar condition applies to the H$\alpha$ observations, where the line profile covers at least three spectral resolution elements across all fibers. This means that, in reality, the number of fully independent data points is $\ge 7$, making a seven-parameter fit viable for H\,{\sc i} data in Arp\,72c. Similarly, for the H$\alpha$ data in Arp\,72b, the number of independent data points is $\ge 8$. Moreover, this confirms that the line is spectrally resolved, thereby justifying the inclusion of velocity dispersion as a fit parameter.

\begin{figure*}[ht]
        \centering
        \includegraphics[width=0.9\hsize]{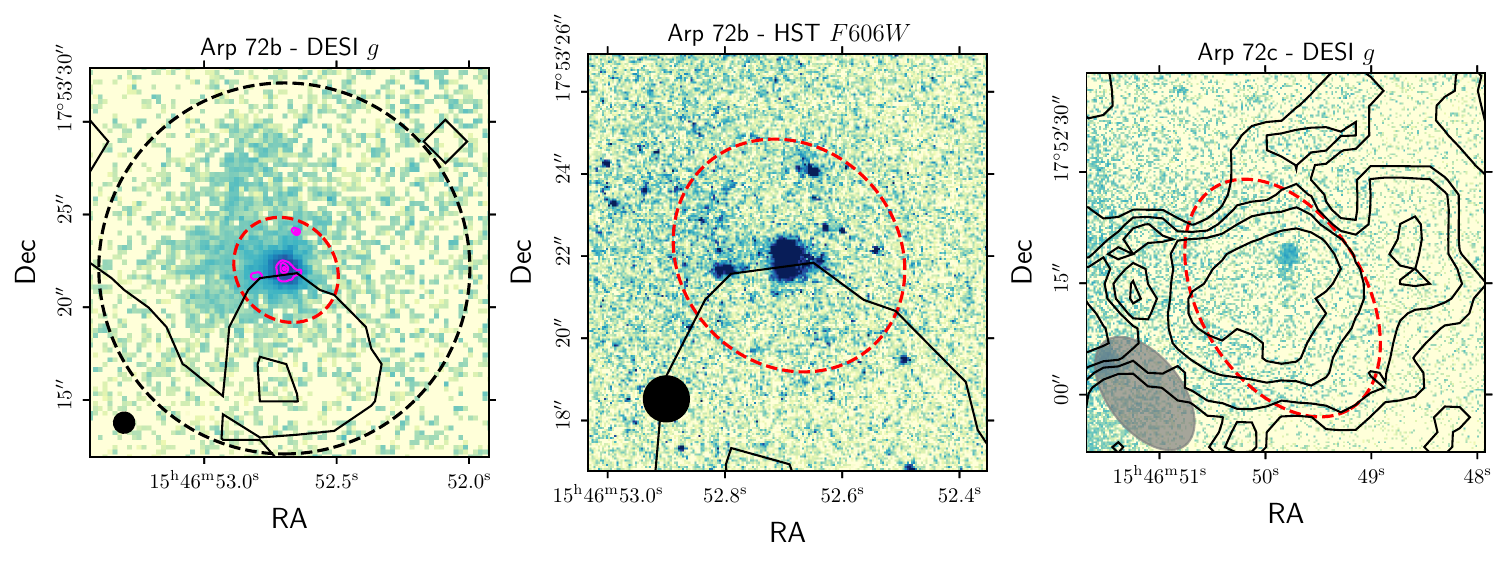}
        \caption{Optical images of the objects overlaid with H\,{\sc i} column density contours (solid black line). The contour levels are 0.6, 3.5, 6.3, 9.1, 11.9, 14.7, and 17.5 $\times 10^{20}$cm$^{-2}$. The optical images were $g$-band from DESI or $F606W$ from HST as indicated above each image. The red dashed ellipse indicates the disk from the 3D fit. In the right panel, the filled gray ellipse represents the GMRT beam. The filled black circle (left and middle panels) indicates the PSF of the GTC data. Left panel: The dashed black circle marks the 10-arcsecond aperture used to search the H\,{\sc i} emission associated with Arp\,72b, and the magenta contours represent the HST $F606W$ emission.}
        \label{all_optical_imaages}
    \end{figure*}

    The disks derived from the 3D fits for each object are shown in Fig. \ref{all_optical_imaages} as red dashed ellipses. Each disk is projected onto the sky according to the geometric parameters listed in Table \ref{all_results-3dfits}, with the disk size determined by $R_{\rm out}$, where $R_{\rm out}$ is the radius up to which we are measuring the dynamical mass. In the same figure, we presented the neutral hydrogen distribution using black column density contours ($N_{{\rm H}\,\text{{\sc i}}}$) overlaid on either the DESI $g$-band or HST $F606W$ images, depending on the object. The region defined by the disk was used to measure $M_{{\rm H}\,\text{{\sc i}}}$. Within this region, we spatially integrated the GMRT H\,{\sc i} spectral cube to obtain the total H\,{\sc i} flux for Arp\,72c, and then calculated $M_{{\rm H}\,\text{{\sc i}}}$ using Eq. \ref{HI_mass}. However, for Arp\,72b, we used the black dashed circular aperture (Fig. \ref{all_optical_imaages}, left panel) to search the associated neutral gas.

    In Figs.~\ref{rotation_dispersion_arp72b_new_version} and \ref{rotation_dispersion_arp72c_new_version} we show the rotation velocity (${v}_{\rm rot}$, upper panels) and the observed velocity dispersion ($\sigma_{v}$, lower panels) as a function of galactocentric radius $R$, for Arp\,72b and Arp\,72c separately. The rotation velocities derived from the observed data for each spaxel or fiber ($M_{1}$ corrected for inclination) are shown as small red squares, while the rotation velocities derived from the reconstructed datacube of the 3D fit are shown as small blue circles. The median values of these points are shown as larger squares and circles. The rotation velocities obtained from the 3D fixed-ring fit are shown as black dots (Arp\,72b - H$\alpha$ fit), while in the case of the parametrized rotation curve method  this is shown with a solid black line (Arp\,72c - H\,{\sc i} fit). In the lower panels of each figure, the green shaded region represents the assumed uncertainty for $\sigma_{v}$. The parameters derived from these fits are listed in Table \ref{all_results-3dfits}. These graphs facilitate a direct comparison of the 3D fit results, as the information extracted from the moment maps ensures that instrumental effects are consistently present in both the data and our best-fit model.

    Figs. \ref{3Dfit_arp72b_new_version} and \ref{3Dfit_arp72c_new_version} present the results of the 3D fit, comparing the moment maps (intensity, velocity, and velocity dispersion) derived from the data with our best-fit model. We also include the SNR map and representative spectra at various positions across the TDG. In the spectral panels, the noise level is indicated by a dashed gray horizontal line, while $v_{\rm sys}$ is marked by a dashed black vertical line. This $v_{\rm sys}$ reference highlights the velocity shifts resulting from the rotation of the TDG. For Arp\,72b (Fig. \ref{3Dfit_arp72b_new_version}), we specifically indicate the rings employed in the fixed-ring method.
    
    When we used the fixed-ring method to obtain an estimate of $R_{\rm d}$ for calculating the dynamical mass, we approximated this parameter by fitting the rotation velocities (black dots in Fig. \ref{rotation_dispersion_arp72b_new_version}, upper panels) to Eq. \ref{vrot}. However, in this case, the final value of this parameter is the average between the fit and the values given by the optical profiles.

    The dynamical mass derived using the parameters in Table \ref{all_results-3dfits} and the measurements of the baryonic mass of each object are presented in Table \ref{Tabla_resultados_general}.

    \begin{table*}[!bt]
    \caption{Parameters resulting from the 3D fits.}
    \centering
    \begin{adjustbox}{width=\textwidth,center}
    
    \begin{tabular}{cccccccccccc} 
    \hline \hline

    Object & \begin{tabular}[c]{@{}c@{}}RA$_{\rm{kc}}$ or $x_{\rm kc}$\\ $\delta_{\rm{kc}}$ or $y_{\rm kc}$\end{tabular} & \begin{tabular}[c]{@{}c@{}}$\varphi_{0}$\\ {[}$^{\circ}${]}\end{tabular} & \begin{tabular}[c]{@{}c@{}}$i$\\ {[}$^{\circ}${]}\end{tabular} & \begin{tabular}[c]{@{}c@{}}$R_{\rm{c}}$\\ {[}kpc{]}\end{tabular}  & \begin{tabular}[c]{@{}c@{}}$v_{\rm{sys}}$\\ {[}km/s{]}\end{tabular} & \begin{tabular}[c]{@{}c@{}} $\Sigma_{0}$ \\ {[}M$_{\odot}$/pc$^{2}${]} \end{tabular} &  \begin{tabular}[c]{@{}c@{}}$R_{\rm{d}}$ \\ {[}kpc{]} \end{tabular} & \begin{tabular}[c]{@{}c@{}}$\sigma_{v}$\\ {[}km/s{]}\end{tabular} & \begin{tabular}[c]{@{}c@{}}$R_{\rm{out}}$\\ {[}kpc{]}\end{tabular} & \begin{tabular}[c]{@{}c@{}}$v_{\rm{rot}}$\\ {[}km/s{]}\end{tabular} & \begin{tabular}[c]{@{}c@{}}$v_{\rm{c}}$\\ {[}km/s{]}\end{tabular}\\ \hline \hline

    Arp\,72b - H$\alpha$ & \begin{tabular}[c]{@{}c@{}}0.5 arcsec\\ 0 arcsec\end{tabular} & 224 $\pm$ 5 & 28 $\pm$ 14 & --  & $3237.5\pm0.1$ & 16 $\pm$ 10\tablefootmark{*} & 0.8 $\pm$ 0.4 & $14.6\pm6.5$ & 0.6 $\pm$ 0.1 & 9.3 $\pm$ 7.5 & $16^{+10}_{-4}$\\

    Arp\,72c - H\,{\sc i} & \begin{tabular}[c]{@{}c@{}} 15:46:49.8 \\ +17:52:13.0 \end{tabular} & 32 $\pm$ 3 & 50 $\pm$ 14 & 2.3 $\pm$ 1.8  & $3303\pm1$ & 7 $\pm$ 12 & 4.6 $\pm$ 1.3 & 14.9 $\pm$ 6.8 & 4.5 $\pm$ 0.3 & 15.2 $\pm$ 7.8\tablefootmark{$\dagger$} & $21^{+9}_{-6}$ \\

    \hline
    \end{tabular}
    \end{adjustbox}
    \tablefoot{Since the H$\alpha$ data could not be astrometrically calibrated, the coordinates of the kinematic center are measured in arcseconds relative to the center of the detector. The error for $R_{\rm c}$ is half the spatial resolution.\\ \tablefoottext{*}{The uncertainty is given by the fit of the rotation velocities to Eq. \ref{vrot}.}
    \tablefoottext{$\dagger$}{With the 3D fit using the parameterized rotation curve, the uncertainty in ${v}_{\rm rot}$ was calculated using the uncertainties of $R_{\rm d}$ and $\Sigma_{0}$; for Arp\,72c - H\,{\sc i}, the distribution of $\Sigma_{0}$ was obtained using a truncated normal distribution to handle negative values with no physical meaning.}}
    \label{all_results-3dfits}
    \end{table*}

\subsection{Arp\texorpdfstring{\,}{ }72b - {\rm H}\texorpdfstring{$\alpha$}{alfa} fit}

    \begin{figure}[!b]
        \centering
        \includegraphics[width=0.9\hsize]{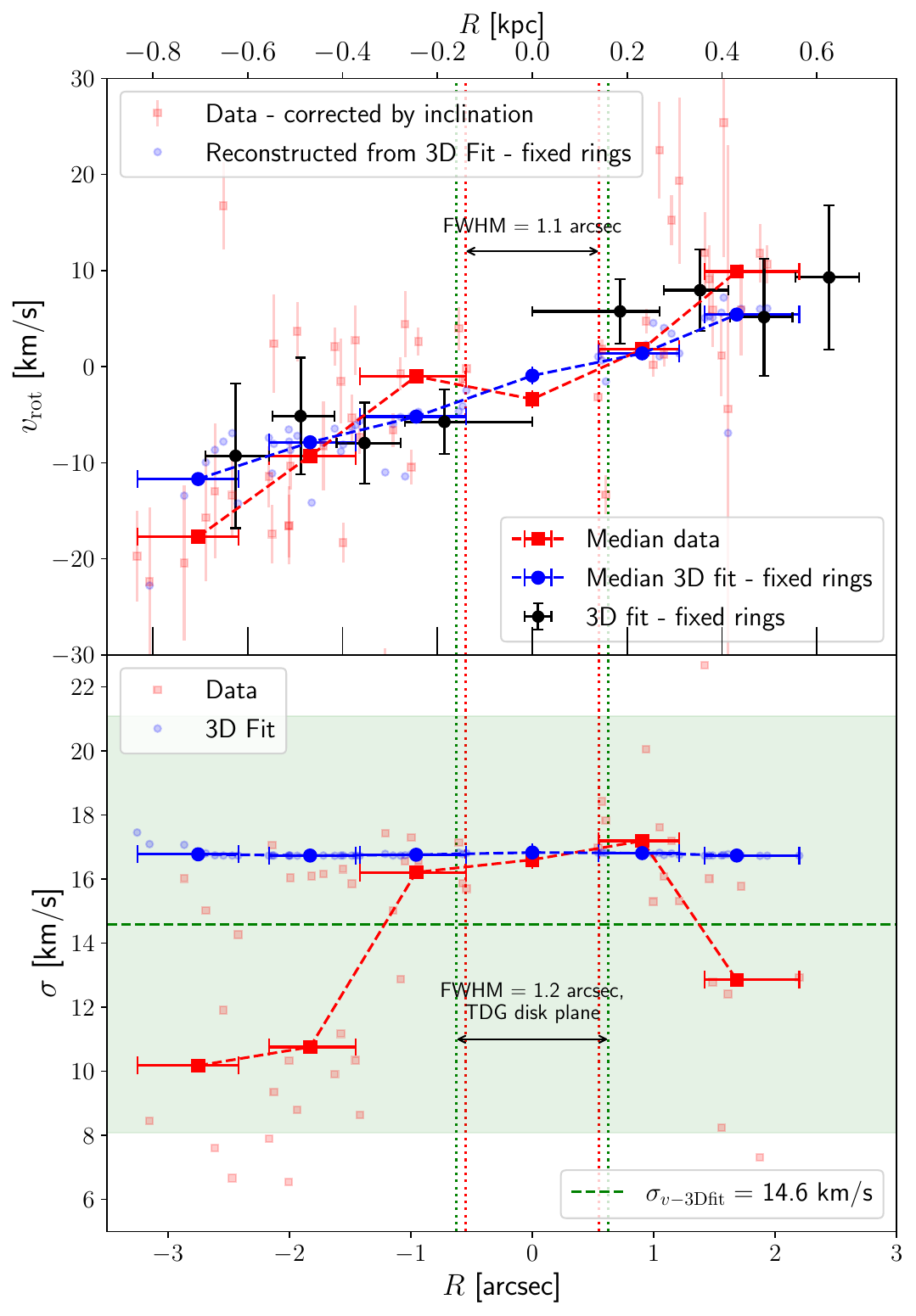}
        \caption{Rotation velocity (${v}_{\rm rot}$, upper panel) and velocity dispersion ($\sigma_{v}$, lower panel) versus galactocentric radius $R$ in Arp\,72b using H$\alpha$ data cube. The small red squares represent the values for all fibers from the observational data whereas the small blue circles indicate the values from the reconstructed cube of the best 3D fit, calculated with equation~\ref{vrot2D} based on their corresponding velocity maps. The large red squares and large blue circles show the median of these points within the intervals indicated by the horizontal error bars. The black circles correspond to the rotation velocity obtained from the 3D fit (fixed rings method). The horizontal dashed green line in the bottom panel marks the value of $\sigma_{v}$ from the fit and the green shaded region represents the assumed uncertainty for $\sigma_{v}$. Blue-shifted and red-shifted regions correspond to negative and positive values of $R$, respectively. The dotted vertical lines represent the spatial resolution and its corresponding projection onto the plane of the disk. The horizontal error bars are determined by the maximum and minimum values of $R$ for the points in each ring.}
        \label{rotation_dispersion_arp72b_new_version}
    \end{figure}
    
    Arp\,72b exhibits brighter H$\alpha$ emission compared to Arp\,72c. The central panel of Fig. \ref{all_optical_imaages} shows the HST image of Arp\,72b in $F606W$ filter, while the left panel shows the same source, but with a larger field of view and the DESI $g$ image. The red ellipse is the same in both pictures, and marks the disk from the 3D fit of the H$\alpha$ data. In the left panel, contours for the HST image are overlaid in magenta on the DESI $g$ image. In both the left and central panels, H\,{\sc i} contours are shown in black.
    
    The absence of any associated H\,{\sc i} cloud and its location outside the tidal arms suggest a relatively low H\,{\sc i} content (see also Fig. \ref{galaxy_desi_HI_contour}). This implies a column density below the GMRT detection threshold, which excludes kinematic analysis with H\,{\sc i} data. The H\,{\sc i} emission (contours in Fig.\,\ref{all_optical_imaages}, middle panel) near the optical center shows a difference of $\sim$100 km/s from the measured $v_{\rm sys}$ with H$\alpha$, and therefore is not associated with Arp\,72b. We therefore applied our methodology to the H$\alpha$ observations instead, with the observed velocity map in Fig. \ref{3Dfit_arp72b_new_version} revealing clear rotational signatures. Arp\,72b presents a particularly interesting case, as it appears largely detached from tidal debris. As one of the oldest systems in Arp\,72 (based on time since the main star formation burst, see Table \ref{stellarmass_age_table}), it should represent the most dynamically independent system and likely has had sufficient time to achieve dynamical equilibrium.

    A notable consideration is that while moment maps are affected by double-peaked H$\alpha$ profiles (since $v_{\rm{obs}}$ represents intensity-weighted velocities), the 3D fit employs a single Gaussian profile along the spectral axis, preferentially weighting the primary peak of the line. At the same time, the 3D fit is limited to the brightest inner region of Arp\,72b where H$\alpha$ emission is detectable, making the kinematics of diffuse emission outside the red ellipse in left and central panels of Fig. \ref{all_optical_imaages} uncertain.
    
    To more accurately replicate the observed intensity map and better account for beam smearing effects, we used only the internal length scale ($R_{\rm{d,int}} = 33$ pc) in the 3D-fit. This is because we assumed that the H$\alpha$ emission is significantly more compact than all the emission detected by the $F606W$ filter. This approach is also based on the sensitivity limits of MEGARA; therefore, the observations are limited to the inner regions, where surface brightness of H$\alpha$ predominates due to star formation. We achieved a consistent fit to all parameters through the fixed rings method. The resulting parameters from this fitting procedure are presented in Table\,\ref{all_results-3dfits}.

\begin{figure}[!ht]
        \centering
        \includegraphics[width=\hsize]{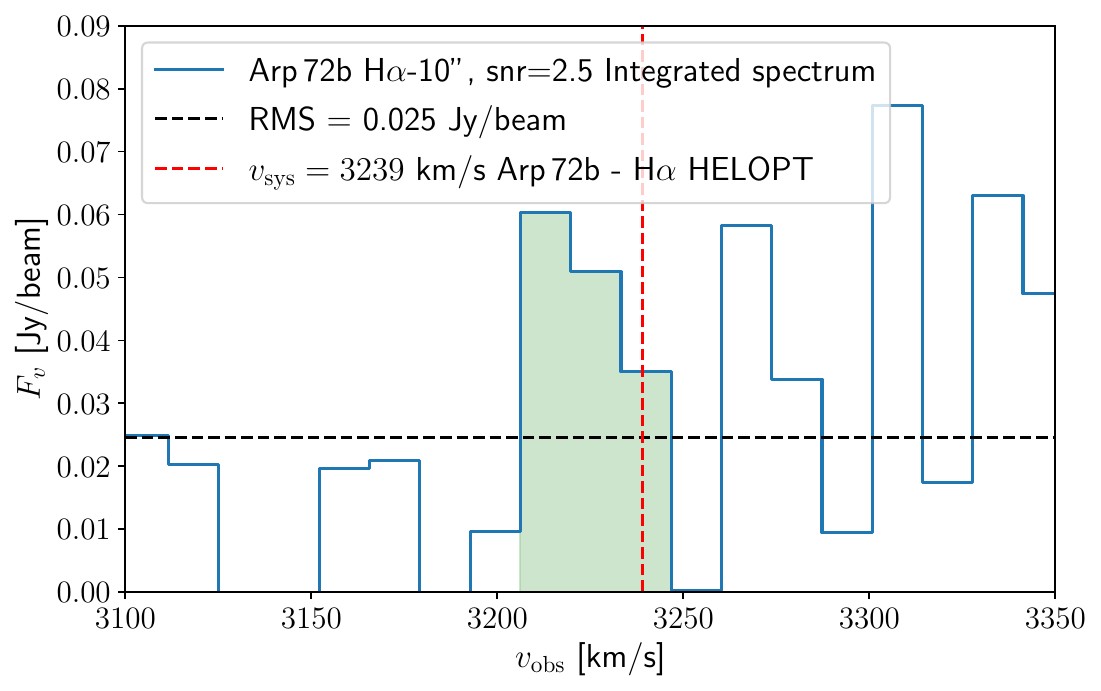}
        \caption{Integrated H\,{\sc i} spectrum of Arp\,72b within the black dashed circular region in Fig. \ref{all_optical_imaages}, central bottom panel. The dashed black horizontal line is the noise level of the spectrum, and the dashed red line is the systemic velocity from 3D-fit corrected to the heliocentric reference frame. Helio-centric velocities are plotted for both the tracers.}
        \label{Arp72b HI spectrum}
    \end{figure}

    The kinematically fitted disk size (red ellipse in Fig. \ref{all_optical_imaages}, left and central panels) defines the radius used for estimating the dynamical mass of the system, corresponding to $R_{\rm{out}}$ in Table \ref{all_results-3dfits}. The diffuse emission of Arp\,72b extends beyond this boundary, making $R_{\rm{out}}$ an underestimate, but as mentioned above, the measurement of the rotation velocity is limited by the noise in the observations. No H\,{\sc i} is detected within this initial radius, which prompts us to extend the integration region in the H\,{\sc i} data cube to the black dashed circle (10 arcsecond radius) in Fig. \ref{all_optical_imaages}, left panel, thus obtaining the spectrum shown in Fig. \ref{Arp72b HI spectrum}. The noise level, which is the rms outside the detection, is marked by the horizontal dashed line. Since the GMRT data are in the heliocentric-optical reference frame, we corrected the $v_{\rm{sys}}$ of Arp\,72b obtained with H$\alpha$ to this same reference system. $v_{\rm{sys}}$ is indicated by the vertical dashed line. This spectrum reveals a weak emission (SNR\,$=$\,2.5) near the systemic velocity from the 3D fit. The H\,{\sc i} gas in the vicinity of Arp\,72b may or may not belong to Arp\,72b. There are uncertainties in both the spectral axis (difference between $v_{\rm sys}$ and the peak of the HI line) and the spatial location of the H\,{\sc i} ($\sim$\,10\,arcsec due to the integration area or an astrometric error associated with interferometric observations of $\sim$\,3.6\,arcsec, see \citealp{annurev:/content/journals/10.1146/annurev-astro-081913-040006}). Despite these uncertainties, we use the HI spectrum to estimate the mass of H\,{\sc i} associated with Arp\,72b and to calculate a lower limit for the gas surface mass density assuming that it is uniform within the 10 arcsec circle.
    
    The high resolution of the HST has enabled us to confirm the morphology derived from the H$\alpha$ surface brightness. The HST image resolves three distinct star-forming regions, including a central and dominant region at the kinematic center, while the other two regions are located within the blue-shifted zones. The red-shifted region aligns with a filamentary structure connected to the central area (magenta countours in Fig. \ref{all_optical_imaages}, left panel and data velocity map in Fig. \ref{3Dfit_arp72b_new_version}).

    Some features of our fit are discussed in Appendix \ref{moments fit Arp72b} and the kinematic results for Arp\,72b obtained with the 3D fit - fixed rings are summarized in Fig. \ref{rotation_dispersion_arp72b_new_version}. This figure allows to evaluate the effects of beam smearing and sampling on the rotation of the system (and similarly on $\sigma_{v}$). The impact of beam smearing and spatial sampling is evident in the $v_{\rm{rot}}$ points of the fit (blue circles, upper panel). For example, the blue point around -2\,arcsec, representing the median $v_{\rm{rot}}$, shows an increase compared to the rotation velocity from the 3D fit (black points). After instrumental effects, $v_{\rm{rot}}$ for that ring increases when obtained from the velocity map, clearly because the spectra at each position are contaminated by nearby areas. In the lower panel of the same figure, we see that for many fibers $\sigma_{v}$ is lower in the data than in the model (horizontal dashed green line). However, the central fibers, where the signal-to-noise ratio is higher, show good agreement. This consistency was further confirmed through a fiber-by-fiber inspection of the fit. The rotation velocity (black points on the top panel) derived from our fit presented in Fig. \ref{rotation_dispersion_arp72b_new_version} is consistent with a disk with $R_{\rm{d}}=0.81$ kpc and $\Sigma_{0}=16.5$ M$_{\odot}$/pc$^{2}$. $R_{\rm{d}}$ is the average value between the external optical length scale (Fig. \ref{F606W_profiles}) and the value obtained by fitting Eq. \ref{vrot} to the rotation velocity. This value is reported in Table \ref{all_results-3dfits}. It should be emphasized that the spectra in each fiber are significantly affected by the PSF and by the sampling used to obtain the spectrum of each fiber, both of which combine information from fibers that are close to each other. The irregularities in the data velocity map (Fig. \ref{3Dfit_arp72b_new_version}) are also visible in some fibers in the upper panel of Fig.\,\ref{rotation_dispersion_arp72b_new_version}. Several fibers in the data (small red squares) deviate significantly from the median $v_{\rm{rot}}$, likely due to wind effects or the low signal-to-noise ratio in certain regions. Nevertheless, the median rotation velocity derived from the observational data still reflects the overall rotation of the system.

\subsection{Arp\texorpdfstring{\,}{ }72c - \texorpdfstring{{\rm H}\,{\sc i}}{HI} fit}

    \begin{figure}[!ht]
        \centering
        \includegraphics[width=\hsize]{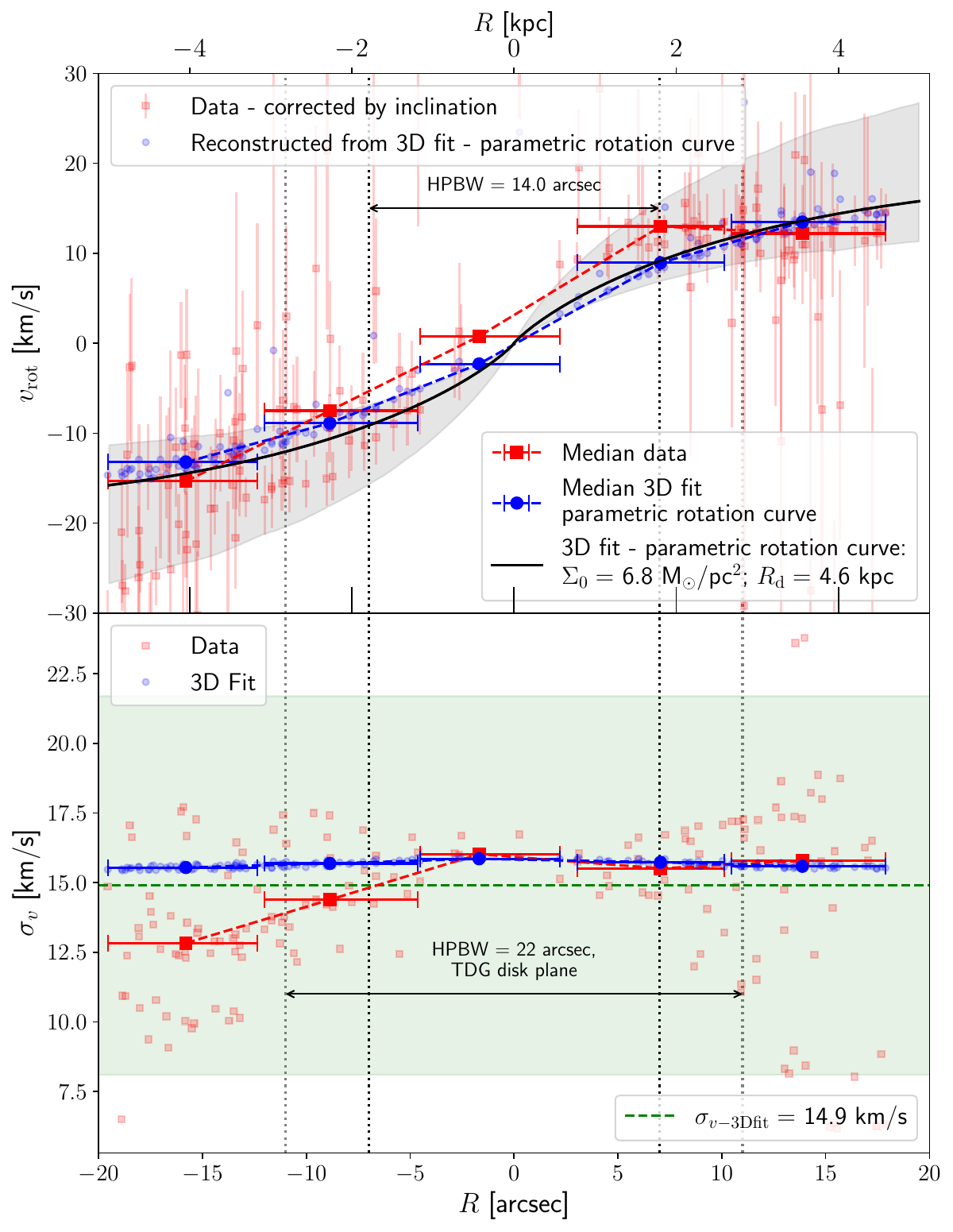}
        \caption{Same description as in Fig. \ref{rotation_dispersion_arp72b_new_version}, but in this case for Arp\,72c using H\,{\sc i} data cube. In the upper panel the rotation curve is given by the solid black line (parametric rotation curve method) and the gray shaded area is the 1$\sigma$ confidence interval of the parameters from 3D fit.}
        \label{rotation_dispersion_arp72c_new_version}
    \end{figure}

    The neutral gas analysis characterizes Arp\,72c as an extended H\,{\sc i} cloud spanning approximately 10 kpc in diameter (see Fig. \ref{all_optical_imaages}, right panel). Arp\,72c-H\,{\sc i} is the largest H\,{\sc i} overdensity in its respective tidal tail, but it is a detached system from the tidal debris. The H\,{\sc i} model for Arp\,72c, described by the parameters in Table \ref{all_results-3dfits} using the parametric rotation curve method, exhibits good consistency with a rotating disk scenario. The quality of the 3D fit is discussed in Appendix \ref{moments and pv Arp72c}, and the dynamics of the neutral gas resulting from this fit is shown in Fig. \ref{rotation_dispersion_arp72c_new_version}. Within the uncertainties the derived values of $v_{\rm{rot}}$ and $\sigma_{v}$ show good agreement between the observational data and our best-fit model (large red squares with large blue dots in both panels of Fig. \ref{rotation_dispersion_arp72c_new_version}). Given that $v_{\rm rot}$ in the data (small red squares in the upper panel) show larger scatter in the blueshifted region, the uncertainties in the rotation curve are higher in this area. The velocity dispersion at large radii tends to be systematically lower than our model predictions, particularly in the blueshifted region.
    
    In order to obtain a value for $R_{\rm{out}}$ or the disk size necessary to estimate the dynamical mass with its respective uncertainty, we perform a 2D fit of the velocity map obtained from the data, fixing the parameters according to the results of the 3D fit. The radius of the last ring is the value we report in Table \ref{all_results-3dfits} as $R_{\rm out}$. 
    
\begin{figure}[ht]
        \centering
        \includegraphics[width=\hsize]{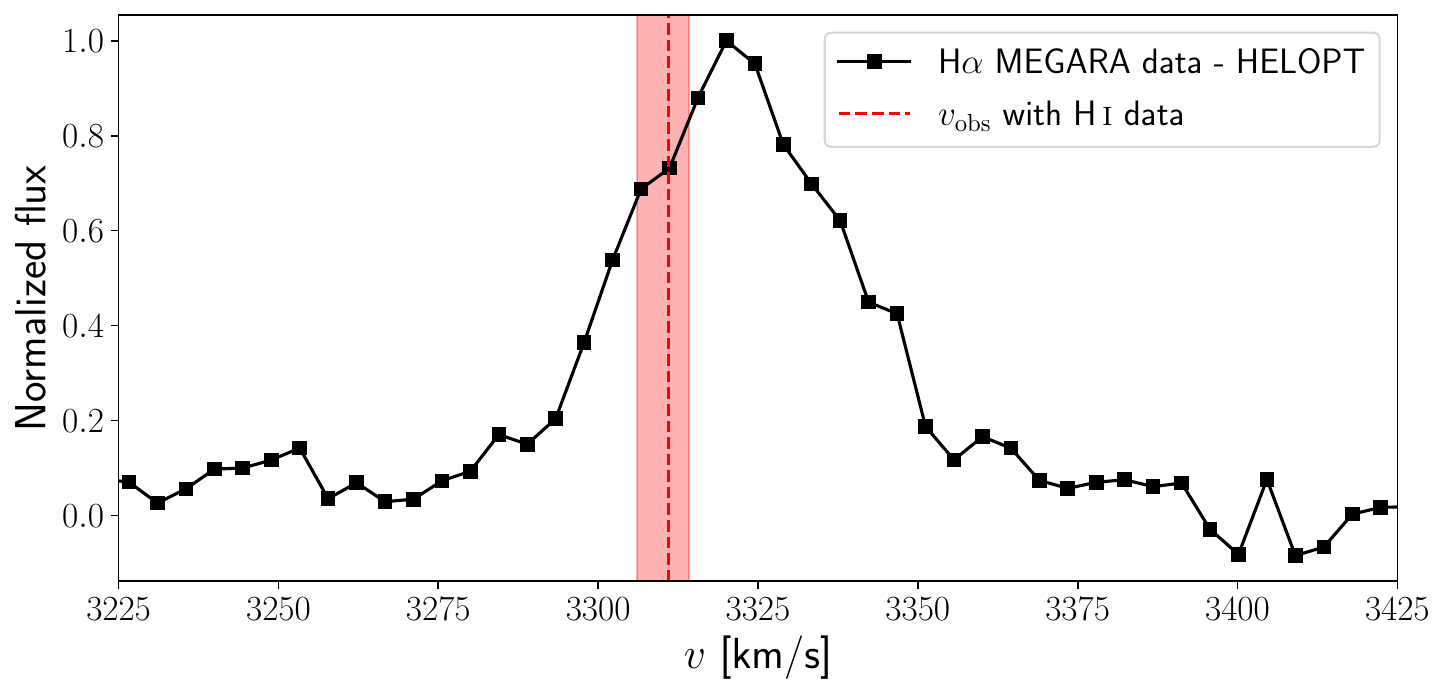}
        \caption{Integrated spectrum of Arp\,72c from MEGARA data, obtained by summing the spectra of the fibers where H$\alpha$ was detected. The dashed vertical line marks the $v_{\rm obs}$ predicted by our best-fit model at the position of the star-forming region of Arp\,72c. The red shaded area represents the uncertainty in $v_{\rm obs}$ due to the size of the GMRT beam. Helio-centric velocities are plotted for both the tracers.}
        \label{Arp72c Halfa integrated spectrum}
    \end{figure}

    Finally, in Fig. \ref{Arp72c Halfa integrated spectrum}, using the MEGARA data for Arp\,72c, we compare the integrated spectrum (in the HELOPT reference frame) over the fibers where H$\alpha$ was detected with the $v_{\rm obs}$ (dashed vertical line) predicted by our disk model at the position of the star-forming region in Arp\,72c. Except for an offset of a few km/s, the star-forming region moves consistently with our disk model for the H\,{\sc i} gas.

\section{Discussion}
\label{Discussion}

    \begin{table}[!ht]
    \caption{Results for the dynamical-to-baryonic mass ratio.}
    \centering
    \begin{tabular}{cccc}
    \hline
    \hline

    Object & $M_{\rm{dyn}}$ {[}$10^{8}$M$_{\odot}${]} & \begin{tabular}[c]{@{}c@{}}$M_{\rm{H}\,\text{\sc{i}}}$ {[}$10^{8}$M$_{\odot}${]}\\ $M_{\rm{gas}}$ {[}$10^{8}$M$_{\odot}${]}\\ $M_{\rm{bary}}$ {[}$10^{8}$M$_{\odot}${]}\end{tabular} & $\dfrac{M_{\rm{dyn}}}{M_{\rm{bary}}}$ \\ \hline \hline

    Arp\,72b - H${\alpha}$ & $0.4^{+0.6}_{-0.2}$ & \begin{tabular}[c]{@{}c@{}}$0.283 \pm 0.090$\\ $0.40 \pm 0.13$\\ $0.57 \pm 0.13$\end{tabular} & $0.7^{+1.2}_{-0.3}$ \\ \hline

    Arp\,72c - H\,{\sc i} & $4.6^{+4.6}_{-2.1}$ & \begin{tabular}[c]{@{}c@{}}$3.04 \pm 0.25$\\ $4.26 \pm 0.35$\\ $4.29 \pm 0.35$\end{tabular} & $1.1^{+1.1}_{-0.5}$ \\ \hline

    \end{tabular}
    \tablefoot{Dynamical mass of the TDGs obtained from the 3D fits and their respective measurements of baryonic mass.}
    \label{Tabla_resultados_general}
    \end{table}

    For agreement with the $\Lambda$-Cold Dark Matter model ($\Lambda$CDM) predictions, the TDGs should consist solely of baryonic matter (gas and stars), resulting in a dynamical-to-baryonic mass ratio ($M_{\rm{dyn}}/M_{\rm{bary}}$) close to unity. However, some TDGs, particularly those displaying disturbed kinematics or recent formation histories, may not have achieved complete dynamical relaxation. The most isolated and evolved TDGs consequently offer the most reliable tests of this theoretical framework. H\,{\sc i} flux measurements are particularly significant as the gas component typically dominates the baryonic mass budget in TDGs ($M_{\rm{bary}}$).
    
    Table \ref{Tabla_resultados_general} gives our estimates of $M_{\rm dyn}$/$M_{\rm bary}$. Several factors contribute to measurement uncertainties. Beyond limitations in our 3D kinematic modeling, the spatial localization of 21 cm line emission derived from interferometric data carries an inherent astrometric error of $\theta_{\rm{therm}} \approx 0.5 \cdot \theta_{\rm{beam}}/\rm{SNR}$ \citep{annurev:/content/journals/10.1146/annurev-astro-081913-040006}. The beam major axis of 18 arcsecond further complicates spectral cube integration in low signal-to-noise regimes, with Arp\,72b representing the most uncertain case. Our analysis for the H$\alpha$ rotation curve of Arp\,72b - H$\alpha$ forces H\,{\sc i} detection near $v_{\rm{sys}}$ since the pixel-by-pixel emission of H\,{\sc i} falls below the sensitivity thresholds. Although spatial integration around the target position yields the spectrum shown in Fig. \ref{Arp72b HI spectrum}, the marginal SNR prevents a definitive association of this gas with Arp\,72b, and in this case, our measurement of the neutral gas gives $M_{\rm dyn}/M_{\rm bary}$ smaller than unity. This is because we may be overestimating $M_{\rm{H}\,\text{\sc{i}}}$ associated with Arp\,72b.

\subsection{Dynamical state and survival of TDGs}

    The study by \cite{10.1093/mnrasl/slv189} examines whether TDGs can be considered virialized rotating disks, thereby validating their positions on the BTFR. Their analysis utilizes H${\alpha}$ kinematics of TDGs previously investigated by \cite{Lelli2015}. The spectral resolution of 21 km/s in their data proves somewhat limiting for TDGs with rotation speeds and velocity dispersions comparable to this resolution (MEGARA data, by comparison, achieves 12.9 km/s). A key argument presented involves the observed offset between the H\,{\sc i} disk center and its optical counterpart, which challenges an idealized disk model. However, this offset may be expected given that TDGs are forming systems with irregular gas distributions. In such cases, the collapse of gas to form stars might not align precisely with the H\,{\sc i} disk center. This scenario presents no contradiction if the H\,{\sc i} disk first reaches equilibrium, followed by the subsequent emergence of star-forming regions that eventually couple to the dynamical center of the system, as exemplified by Arp\,72c. Furthermore, H${\alpha}$ kinematics often exhibit localized perturbations caused by non-gravitational physical processes. This effect likely explains the disturbed velocity fields reported by \cite{10.1093/mnrasl/slv189}, as well as those observed in our H${\alpha}$ data for Arp\,72b. Such perturbations do not necessarily invalidate the overall equilibrium state of these systems, but rather reflect the complex interplay of dynamical and star-forming processes in young TDGs.

    Alternative kinematic tracers such as CO, which like H$\alpha$ is associated with star-forming regions, provide additional means to investigate TDG dynamics. The study by \cite{Querejeta2021} demonstrates this approach, resolving giant molecular clouds (GMCs) within a TDG in the interacting system Arp 94 \citep{10.1093/mnras/277.2.641, Mundell_2004}. Their analysis reveals that compact molecular emission appears more irregular than its diffuse counterpart, suggesting that star-forming regions trace local interactions between GMCs and young star clusters, while the diffuse component reflects the global potential of the TDG. This distinction is analogous to the morphological/kinematic discrepancies often found between H\,{\sc i} and H$\alpha$ tracers. Notably, \cite{Querejeta2021} reports molecular gas masses comparable to atomic gas masses within the TDG. If this is a general characteristic of TDGs, our baryonic mass measurements are underestimated by a factor of $\sim$2.

    The PV diagrams conclusively exclude projection effects for Arp\,72c, demonstrating no significant line-of-sight velocity components distinct from the main kinematic structure. While star formation processes may lead to partial gas loss during the initial phases of TDG formation, high-resolution numerical simulations from \cite{10.1093/mnras/stt2211} indicate that less than 50\% of the gas is typically lost, regardless of the IMF or star formation history. These simulations demonstrate that TDGs can survive their primary star formation bursts and convert only a minor fraction of their gas mass into stars. \cite{10.1093/mnras/stu2629} examine TDG survival within the tidal fields of their host galaxies. Both compressive and stretching tides operate on the TDG. Stretching in the direction of the barycentre of the host galaxies causes mass loss, but compression in the perpendicular direction makes the TDG more compact and enhances the SFR. Their results show that TDGs can persist for more than 3 Gyr, though they ultimately evolve into more compact systems than their original gas clouds.

    These considerations suggest that Arp\,72b and Arp\,72c are at different evolutionary stages. While Arp\,72c is subject to tidal forces, its detached nature allows self-gravity to dominate its dynamics. Consequently, the combined effects of stellar feedback and tidal interactions will likely lead to a more compact configuration over time. Arp\,72b presents a particularly interesting case, exhibiting a compact stellar component without detected associated neutral gas. This morphology could result from either intense star formation efficiently expelling its gas reservoir, or alternatively, from extended tidal interactions given its likely older age (see Table \ref{stellarmass_age_table}).
    
    \cite{Recchi2007} compared the early evolution of isolated dSphs with the early stages of a TDG. They found that the presence of a dark matter halo is not necessary to ensure the survival of a galaxy in the face of star formation bursts, because in models without Dark\,Matter\,(DM), the dynamics of the gas is dominated by its own density and distribution rather than by the depth of the potential well, resulting in a behavior similar to DM-dominated models. The TDGs that would come closest to these conditions are those that are more detached. The objects in this study are promising cases because Arp\,72b is outside detectable tidal tails, and Arp\,72c shows a velocity field of H\,{\sc i} consistent with a rotating disk and is almost at the end of its respective tidal tail.

\subsection{Baryonic and dynamical mass}

    \begin{figure*}[ht]
        \centering
        \includegraphics[width=\hsize]{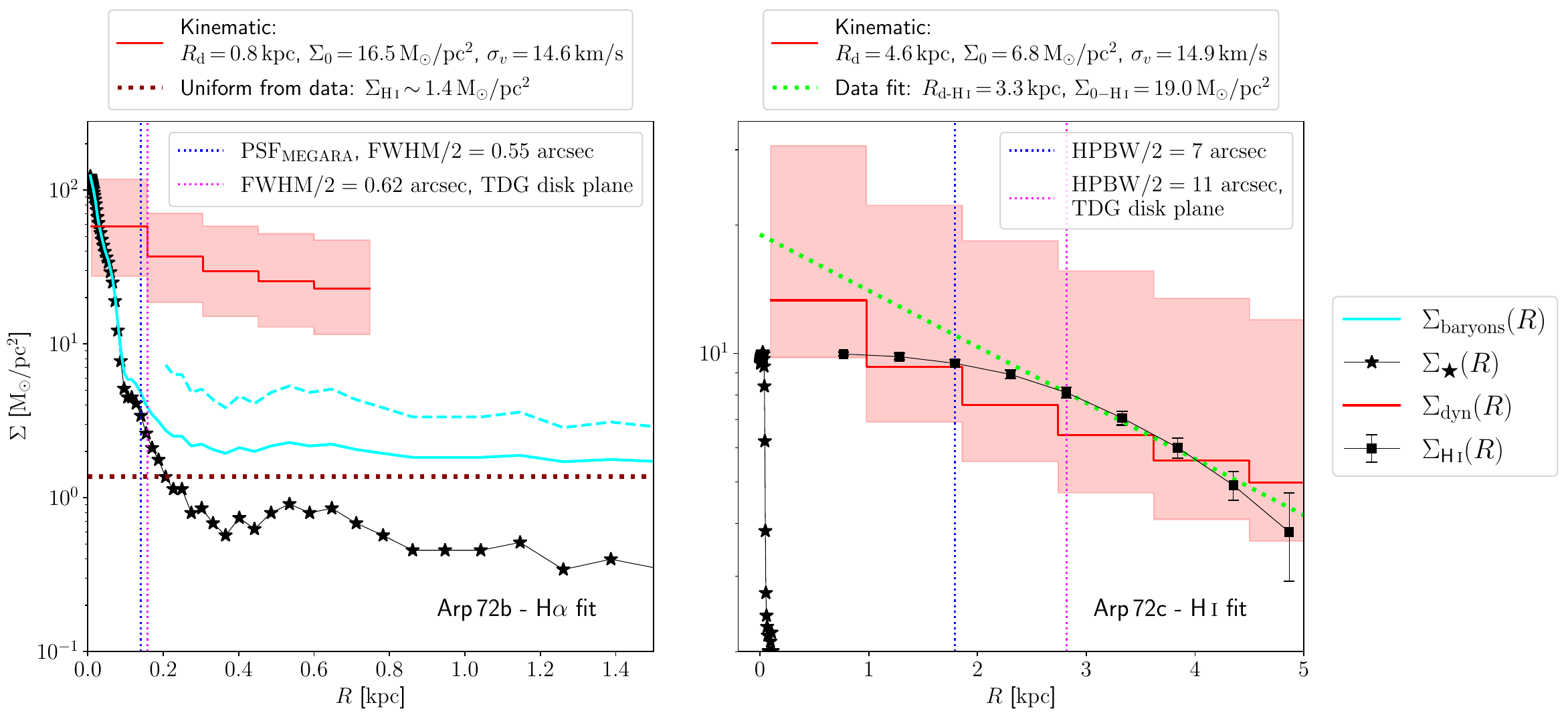}
        \caption{Surface mass densities as a function of the galactocentric radius of the TDGs based on the H\,{\sc i} spectral cube and optical images, together with the mass profile derived from the 3D fit. The red line, with a $1\sigma$ confidence interval (the shaded red region), represents the dynamical mass profile inferred from the kinematics of the object and Eqs. \ref{dynamical_mass} and \ref{vrot}, using rings of equal width. For the profile obtained from optical data (black stars connected by lines) the geometric parameters are those obtained with the {\sc ellipse} task in {\sc iraf}. The blue and magenta (in the plane of the disk) dotted vertical lines marks the threshold beyond which spatially independent data points exist, in the right panel indicates half of the mean beam size (the average of $B_{\rm min}$ and $B_{\rm maj}$), in the left panel denote half of the PSF FWHM for MEGARA-GTC. Left panel: Arp\,72b profiles, the continuous cyan line is the total baryonic surface mass density profile obtained from adding the uniform estimation of $\Sigma_{\rm{H}\,\text{\sc{i}}}$ with the stellar mass profile obtained from the optical profile using the mass-luminosity relations of \cite{Bruzual2003} with an initial mass function (IMF) of Kroupa. The dashed cyan line is $\Sigma_{\rm baryons}$ if the stellar population has an age of 5 Gyr. The dotted dark red line is the uniform estimation of $\Sigma_{\rm{H}\,\text{\sc{i}}}$. Right panel: Arp\,72c profiles, the H\,{\sc i} and stellar profiles were not summed because they are not concentric. The dotted green line is an exponential fit to the outer part of the H\,{\sc i} profile. The H\,{\sc i} profile (black squares connected by lines) was constructed using the geometric parameters of the 3D fit.}
        \label{all_profiles_mass.}
    \end{figure*}

    In order to evaluate our results, we compared the mass distributions derived from rotation curves with those obtained from H\,{\sc i} and optical surface brightness profiles; with $F606W$-HST for Arp\,72b, and primarily with H\,{\sc i} for Arp\,72c. We have used the theoretical mass-luminosity relations obtained by \cite{Bruzual2003} using the metallicities and ages closest to those obtained by \cite{refId0_Javier2024} for each object. These comparisons are presented in Fig.\,\ref{all_profiles_mass.}, where for each object we show, on the one hand, the mass profiles derived from the H\,{\sc i} intensity maps and the optical images, as well as the mass profile estimated with our 3D fit. From Eqs. \ref{dynamical_mass} and \ref{vrot}, we construct a dynamical mass profile to compare with the stellar mass and gas mass distributions. $\Sigma_{\rm dyn}$ complements the mass profile of the thin disk (${v}_{\rm rot}$) with the material traced by the velocity dispersion, thus giving us an estimate of the mass needed to explain the kinematics of the system.
    
    This comparison allows us to evaluate the baryonic mass distribution of each object disregarding the modest differences in $\varphi_{0}$ and $i$ between those obtained using the 3D fit and those obtained using the {\sc ellipse} task in {\sc iraf} (Arp\,72b: $\Delta i$\,$\sim$\,$15^{\circ}$, $\Delta \varphi_{0}$\,$\sim$\,$19^{\circ}$).
    
    For Arp\,72b (Fig. \ref{all_profiles_mass.}, left panel), where no H\,{\sc i} is detected, the dynamical mass (red line) exceeds the baryonic surface density at $R>0.1$ kpc. Although the extended optical profile shows a slope somewhat similar to the dynamical mass profile, an additional mass component is required to explain the measured rotation and dispersion. As a conservative estimate, we assumed a uniform H\,{\sc i} distribution within 10 arcseconds (dotted line in Fig. \ref{all_profiles_mass.}, left panel), but the kinematics implies the presence of additional extended material, likely including both molecular gas and more neutral hydrogen. The presence of a significant amount of H\,{\sc i} is perhaps not likely because it should have been detected by the GMRT observations. However, in Fig.\,\,\ref{all_optical_imaages}, left panel and the stellar mass profile, we see material outside the disk of our 3D fit. At $R_{\rm out}$ we overestimate $\sigma_{v}$ (lower panel, Fig. \ref{rotation_dispersion_arp72b_new_version}); these results both indicate that $\sigma_{v}$ is being affected by stellar feedback. Therefore, $\Sigma_{\rm dyn}$ increases significantly with respect to $\Sigma_{\rm baryons}$. While our H\,{\sc i} estimation yields $M_{\rm dyn}$$\sim$$M_{\rm bary}$, the results indicate that a molecular gas component, combined with an ionized gas contribution, is required to explain the extended dynamics of Arp\,72b if no dark matter halo exists. This is because, even after subtracting the $\sigma_{v}$ contribution from $\Sigma_{\rm dyn}$ at $R>0.2$\,kpc, we find that $\Sigma_{\rm dyn}>\Sigma_{\rm baryons}$. The discrepancies between $\Sigma_{\rm baryons}$ and $\Sigma_{\rm dyn}$ may decrease if the diffuse emission of Arp\,72b belongs to an older stellar population, considering the stellar debris captured during its formation. Therefore, we applied a mass-to-light ratio for a 5\,Gyr stellar population, resulting in the dashed cyan line in the left panel of Fig. \ref{all_profiles_mass.}; while this profile approaches $\Sigma_{\rm dyn}$, it remains insufficient to explain the difference.
    
    Using the rotation curve of Arp\,72c derived from the H\,{\sc i}, our results indicate that $\Sigma_{\rm dyn}$ is highly consistent with $\Sigma_{\rm{H}\,\text{\sc{i}}}$ (Fig. \ref{all_profiles_mass.}, right panel). Consequently, our rotation fit is robust, and the asymmetric drift correction (Eq. \ref{asym_drift}) successfully reproduces the neutral gas distribution. It should also be noted that in Fig. \ref{all_profiles_mass.}, $\Sigma_{\rm dyn}$ is limited by the spatial resolution (dotted vertical lines); therefore, the results are uncertain within this limit. Furthermore, we emphasize that, as in the previous cases, the contribution of molecular gas to the total baryonic mass profile has not been taken into account. The stellar mass distribution (stars connected by lines) is shown separately in this figure because it is not concentric with the gas.

    \begin{figure}[!ht]
        \centering
        \includegraphics[width=\hsize]{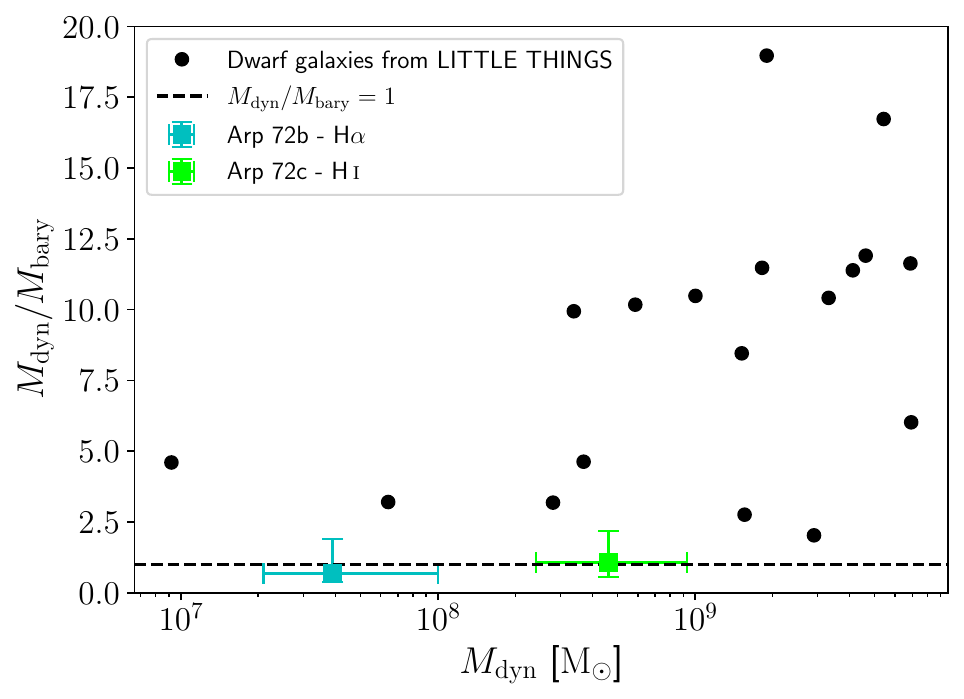}
        \caption{Ratio of dynamical mass and baryonic mass versus the dynamical mass of the TDGs of this study compared with some dwarf galaxies from the LITTLE THINGS sample \citep{2015AJ....149..180O}.}
        \label{mdyn}
    \end{figure}

\subsection{Baryonic Tully-Fisher relation}

    Although TDGs can deviate from equilibrium due to their recent formation and tidal influences, potentially affecting the reliability of both their dynamical-to-baryonic mass ratios, $M_{\rm dyn}/M_{\rm bary}$, and positions on the BTFR, these parameters are still useful for identifying TDGs. In Fig. \ref{mdyn}, we show the ratio between the dynamical mass and baryonic mass as a function of the dynamical mass for the two TDGs, and compare it with that of dwarf galaxies from the LITTLE THINGS sample \citep{2015AJ....149..180O}. Our two TDGs have lower ratios than the dwarf galaxies and within the uncertainties, they are consistent with $M_{\rm dyn}/M_{\rm bary}=1$.
    
\begin{figure}[!ht]
        \centering
        \includegraphics[width=\hsize]{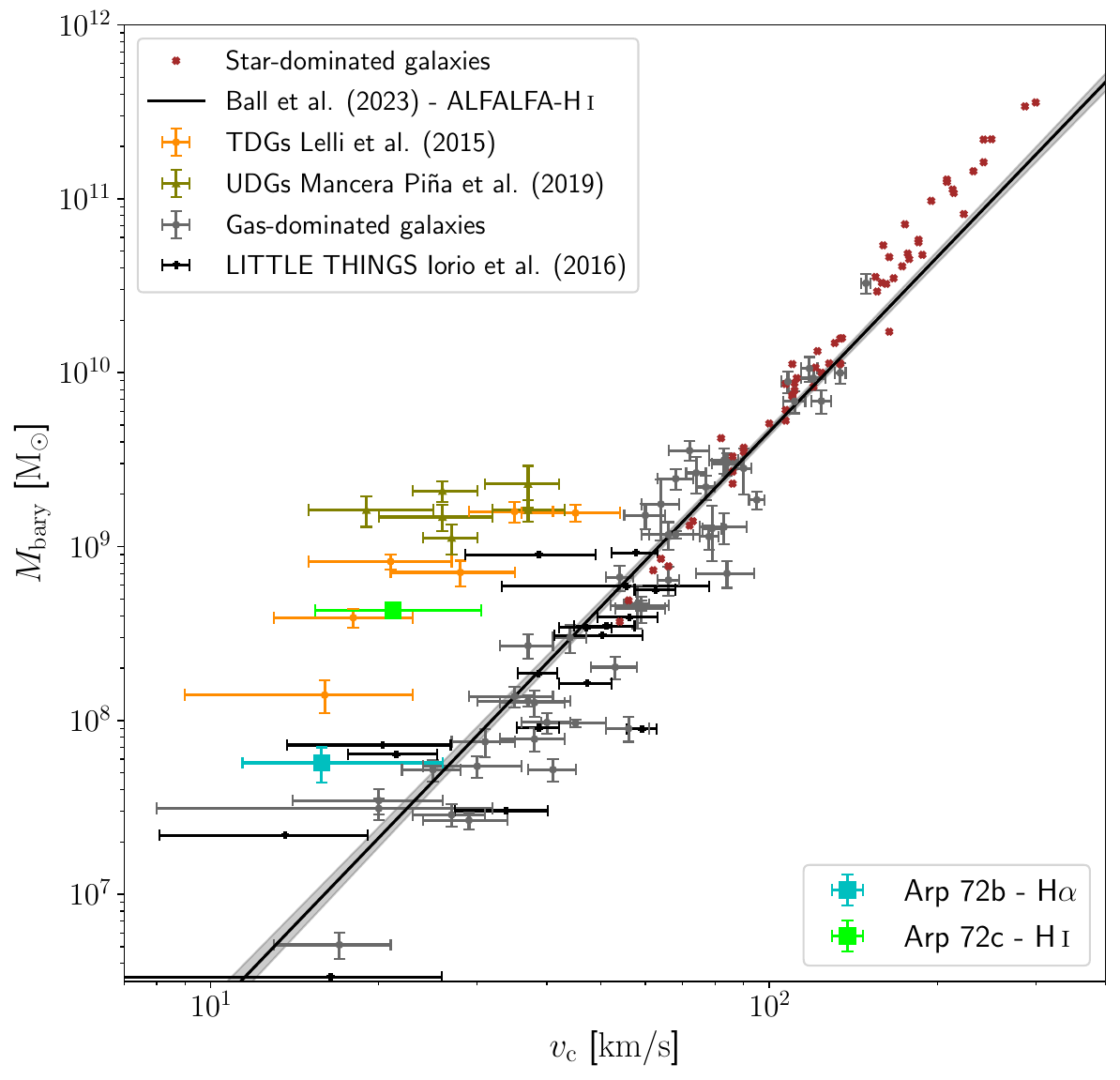}
        \caption{Baryonic mass as a function of circular velocity (BTFR). The objects in our study are plotted alongside the TDGs from \cite{Lelli2015}, UDGs from \cite{ManceraPina2019} and dwarf irregular galaxies from \cite{10.1093/mnras/stw3285}. $v_{\rm{c}}$ is the circular velocity corrected for pressure support (Eq. \ref{asym_drift}). For star-dominated galaxies \citep{2005ApJ...632..859M} and for gas-dominated galaxies \citep{McGaugh_2012} $v_{\rm{c}}$ is the rotation velocity at the flat part of the rotation curve. We show the recent fit for galaxies from Arecibo Legacy Fast ALFA (Arecibo $L$-band Feed Array) Survey (ALFALFA) as a black line \citep{2023ApJ...950...87B}.}
        \label{btfr}
    \end{figure}

    In Fig. \ref{btfr}, the total baryonic mass is plotted against the circular velocity (Eq. \ref{asym_drift}) to evaluate our results concerning the BTFR. We compare our TDGs to gas-dominated and star-dominated galaxies from \cite{2005ApJ...632..859M, McGaugh_2012}, which correlate quite well. We also include the TDGs from \cite{Lelli2015}, studied with the same exponential disk methodology, and the UDGs from \cite{ManceraPina2019}, for which low amounts of dark matter have been measured. These latter two cases separate notably from the BTFR. Our results for Arp\,72c also differ from conventional galaxies (see Fig. \ref{btfr}). While the distinction is less evident for Arp\,72b, its position tends to follow the behavior typical of dark matter-deficient systems.

    The \cite{Lelli2015} TDGs sample includes two TDGs associated with NGC\,7252, one associated with NGC\,4694, and three associated with NGC\,5291. The three NGC\,5291 TDGs were earlier studied by \cite{2007Sci...316.1166B}, who reached quite different conclusions than \cite{Lelli2015}. While \cite{2007Sci...316.1166B} concluded that a considerable amount of unseen matter was present in these TDGs, \cite{Lelli2015} found that the observed baryonic matter could account for all of the dynamical mass. \cite{Lelli2015} used new H\,{\sc i} data, different assumptions about the geometry of the system, and an improved treatment of beam-smearing effects to derive smaller rotational velocities and larger H\,{\sc i} masses than \cite{2007Sci...316.1166B} for these TDGs. \cite{10.1093/mnrasl/slv189} then reanalyzed these three TDGs using new high resolution optical imaging spectroscopy. Using H$\alpha$ velocity fields, they conclude that these TDGs are not relaxed or virialized, and have complex kinematics. Two of the TDGs may lie above the BTFR, though with large uncertainties, while the third appears to fall on the BTFR. This discussion of NGC\,5291 emphasizes that early-stage TDGs are not ideal targets for rotation curve studies of dark matter content. Detached TDGs such as Arp\,72b and Arp\,72c are better targets for such studies.

    There has been considerable discussion in the literature about the cause of the BTFR, its exact mathematical form, its scatter, and the best way to measure the circular velocity that is plotted. The BTFR is a relation between the dynamical and baryonic matter of galaxies, thus it is a function of how a dark matter halo and its baryonic content have evolved with time \citep{McGaugh_2012}. The position of TDGs on the BTFR can change during their evolution due to mass loss and changes in their internal dynamics. For instance, numerical simulations by \cite{10.1093/mnras/stu2629} demonstrate that TDGs evolving in tidal fields develop increasingly steep rotation curves at small radii as they lose mass and grow more compact. However, at larger radii, the rotation curves of their model TDGs are falling because of the lack of dark matter. The rotation curves for our TDGs are rising, perhaps because we are not able to observe at large enough radii to measure a turn-over in the rotation curve. In comparing TDGs to the BTFR, the radius at which the velocity is measured is important.

\section{Conclusions}
\label{Conclusions}

    The H\,{\sc i} data from GMRT and H$\alpha$ data from MEGARA allowed us to resolve the internal kinematics of the TDGs studied in this work, which, in turn, enabled the application of a 3D fitting methodology. Given the complexity of forming low-mass systems such as TDGs, this methodology eliminates instrumental effects on ${v}_{\rm rot}$ and $\sigma_{v}$, while simultaneously allowing for the precise measurement of these parameters for better tracing of the gravitational potential. Although our results apply the same model to objects that may be in different evolutionary stages or have different origins, the fits enable us to evaluate the dynamic state of these objects and determine whether the dynamical mass significantly exceeds the baryonic mass. If such a significant excess is observed, a purely tidal origin for a galaxy appears unlikely. Our analysis using H\,{\sc i} and H$\alpha$ data also allows us to compare these kinematics tracers. The kinematics observed with H$\alpha$ are often more disturbed by star formation effects while simultaneously tracing more compact regions that can respond to local potentials. In contrast, H\,{\sc i} is much more extended, allowing it to trace the global potential of the system, but its utility is limited by the spatial resolution of the data.

    Based on our results and those of \cite{refId0_Javier2024}, the most detached TDGs, Arp\,72b and Arp\,72c, exhibit properties consistent with theoretical expectations for tidal objects. Although further investigation is required to determine the nebular oxygen abundance in Arp\,72c and to obtain more precise stellar metallicity measurements for Arp\,72b, the kinematic data present compelling evidence about the lack of dark matter in these systems. The H$\alpha$ kinematics in Arp\,72b and the H\,{\sc i} kinematics in Arp\,72c both indicate rotating disk structures where velocity dispersion contributes significantly to tracing the gravitational potential. These findings suggest that well-detached tidal dwarf galaxies provide the most reliable systems for testing different formation scenarios of TDGs. However, in Arp\,72b, the lack of detected H\,{\sc i}, the absence of CO observations, and stellar feedback make it difficult for our kinematic results to be fully consistent with the observed baryonic mass.

    Arp\,72b and Arp\,72c are located outside the BTFR, as found by \cite{Lelli2015} for other TDGs and for low dark matter content UDGs studied by \cite{ManceraPina2019}.

    We have identified two objects at distinct evolutionary stages within the Arp\,72 interacting system. Arp\,72c is a gas-rich system that shows kinematics consistent with a rotating disk. In contrast, Arp\,72b appears to be more evolved than Arp\,72c, having lost a significant portion of its gas and more detached from the surrounding tidal debris.

\begin{acknowledgements}
      J.Z-C and O.M.P-N acknowledge support of Grant PID2023-147386NB-I00 funded by MICIU/AEI/10.13039/501100011033 and by ERDF/EU. O.M.P-N. would like to thank the Secretaría de Ciencia, Humanidades, Tecnología e Innovación (SECIHTI) for funding a scholarship, and the Instituto Nacional de Astrofísica, Óptica y Electrónica (INAOE) for providing the facilities and support that made this work possible. G.N.O.L. acknowledges the financial support provided by SECIHTI through grant CBF-2025-I-201. S.C. acknowledges funding from the State Research Agency (AEI-MICIU) of the Spanish Ministry of Science, Innovation, and Universities under the grant `The relic galaxy NGC 1277 as a key to understanding massive galaxies at cosmic noon’ with reference PID2023-149139NB-I00. We thank the staff of the GMRT that made these observations possible. GMRT is run by the National Centre for Radio Astrophysics of the Tata Institute of Fundamental Research. This publication is based on data obtained with the MEGARA instrument at the Gran Telescopio CANARIAS. MEGARA has been built by a Consortium led by the Universidad Complutense de Madrid (Spain) and that also includes the Instituto de Astrofísica, Óptica y Electrónica (Mexico), Instituto de Astrofísica de Andalucia (CSIC, Spain) and the Univesidad Politécnica de Madrid (Spain). MEGARA is funded by the Consortium institutions and by GRANTECAN S.A. Based on observations made with the Gran Telescopio Canarias (GTC), installed in the Spanish Observatorio del Roque de los Muchachos of the Instituto de Astrofisica de Canarias, in the island of La Palma. This work is based on data obtained with MEGARA instrument, funded by European Regional Development Funds (ERDF), through Programa Operativo Canarias FEDER 2014-2020. This research used data obtained with the Dark Energy Spectroscopic Instrument (DESI). DESI construction and operations is managed by the Lawrence Berkeley National Laboratory. This material is based upon work supported by the U.S. Department of Energy, Office of Science, Office of High-Energy Physics, under Contract No. DE–AC02–05CH11231, and by the National Energy Research Scientific Computing Center, a DOE Office of Science User Facility under the same contract. Additional support for DESI was provided by the U.S. National Science Foundation (NSF), Division of Astronomical Sciences under Contract No. AST-0950945 to the NSF's National Optical-Infrared Astronomy Research Laboratory; the Science and Technology Facilities Council of the United Kingdom; the Gordon and Betty Moore Foundation; the Heising-Simons Foundation; the French Alternative Energies and Atomic Energy Commission (CEA); the National Council of Humanities, Science and Technology of Mexico (CONAHCYT); the Ministry of Science and Innovation of Spain (MICINN), and by the DESI Member Institutions: www.desi.lbl.gov/collaborating-institutions. The DESI collaboration is honored to be permitted to conduct scientific research on I'oligam Du'ag (Kitt Peak), a mountain with particular significance to the Tohono O'odham Nation. Any opinions, findings, and conclusions or recommendations expressed in this material are those of the author(s) and do not necessarily reflect the views of the U.S. National Science Foundation, the U.S. Department of Energy, or any of the listed funding agencies. This research is based on observations made with the NASA/ESA Hubble Space Telescope obtained from the Space Telescope Science Institute, which is operated by the Association of Universities for Research in Astronomy, Inc., under NASA contract NAS 5–26555. These observations are associated with program(s) 15446 - Dalcanton, Julianne - University of Washington Establishing HST's Low Redshift Archive of Interacting Systems.
\end{acknowledgements}

\FloatBarrier
\bibliographystyle{aa}
\bibliography{referenceTDGpaper}

\begin{appendix}

\section{Dynamical modeling and uncertainties determination}
\subsection{Dynamical model}
\label{Dynamical model append}
    The velocity field of a disk galaxy with an exponential mass profile follows the rotation curve derived by \cite{1970ApJ...160..811F}:
    \begin{equation}
        {v}_{{\rm rot}}^{2} = 4 \pi {G} \Sigma_{0} R_{\rm{d}} y^{2} \left [ {\rm I}_{0}(y) {\rm K}_{0}(y) - {\rm I}_{1}(y) {\rm K}_{1}(y) \right ] \, ,
        \label{vrot}
    \end{equation}
    \noindent where $\Sigma_{0}$ is the surface mass density at the center of the disk, $R_{\rm{d}}$ is the scale length, $\rm{I}$ and $\rm{K}$ are the modified Bessel functions and $y=R/2R_{\rm{d}}$. The mass within radius $R$ in an exponential disk is:
    \begin{equation}
        M(R) = 2\pi \Sigma_{0} R_{\rm{d}}^{2} \left [1 - \exp{(-R/R_{\rm{d}})} \left (1+\frac{R}{R_{\rm{d}}} \right) \right] \, ,
        \label{expdisc}
    \end{equation}
    \noindent In the absence of random gas motions, the rotational velocity ($v_{\rm{rot}}$) of the system is equal to the circular velocity ($v_{\rm{c}}$), so combining Eqs. (\ref{vrot}) and (\ref{expdisc}) yields the dynamical mass expressed in terms of the $v_{\rm{c}}$:
    \begin{equation}
        \begin{split}
            & M_{\rm{dyn}}(R) = \varepsilon  \dfrac{R v_{\rm{c}}^{2}}{\rm{G}} \, , \\
            & \text{where} \quad \varepsilon = \dfrac{1}{4y^{3}} \dfrac{\left [1 - \exp{(-2y)} \left (1+2y \right) \right]}{{\rm{I}}_{0}(y){\rm{K}}_{0}(y) - {\rm{I}}_{1}(y){\rm{K}}_{1}(y)} \, .
        \end{split}
        \label{dynamical_mass}
    \end{equation}
    However, in dwarf galaxies the velocity dispersion can become comparable to the rotational velocity, therefore, the rotational velocity must be corrected by the asymmetric-drift to obtain $v_{\rm{c}}$. This means, that in this case, $v_{\rm rot}$ is not a direct tracer of the dynamical mass of the system \citep{10.1093/mnras/stw3285}. For the case of a system with isotropic velocity dispersion \citep{binney2011galactic}:
    \begin{equation}
        {v}_{{\rm c}}^{2} = {v}_{{\rm rot}}^{2} - \sigma_{v}^{2} \left (  \frac{\partial \ln (\rho)}{\partial \ln (R)} + \frac{\partial \ln (\sigma_{v}^{2})}{\partial \ln (R)} \right ) \, ,
    \end{equation}
    \noindent where $\rho$ is the density of the disk and $\sigma_{v}$ is the velocity dispersion. If the galaxy scale height is constant with radius, one can obtain $\partial \ln (\rho) / \partial \ln (R)$$=$$ \partial \ln (\Sigma) / \partial \ln (R)$, which for an exponential disk gives $\partial \ln (\Sigma) / \partial \ln (R) $$=$$ - R / R_{\rm{d}}$. If the velocity dispersion is uniform throughout the disk, the asymmetric-drift correction can be written as \citep{1996AJ....111.1551M, 2014A&A...566A..71L}:
    \begin{equation}
        {v}_{{\rm c}}^{2} = {v}_{{\rm rot}}^{2} + \sigma_{v}^{2} \left ( \frac{R}{R_{{\rm d}}} \right ) \, ,
        \label{asym_drift}
    \end{equation}
    \noindent where $R_{\rm{d}}$ is related to the total mass distribution and consequently does not trace the surface brightness distribution of either the 21 cm H\,{\sc i} line or H$\alpha$ emission, which we employ as kinematic tracers. Therefore, we utilize the characteristic scale length $R_{\rm{d}}$ and $R_{\rm{c}}$ (cut radius) for the H\,{\sc i} data to make an approximation of the surface brightness distribution of H\,{\sc i}, $\mu_{\rm{H}\,\text{\sc{i}}}(R)$, such that:
    \begin{equation}
        \mu_{\rm{H}\,\text{\sc{i}}} (R) \propto \left\{\begin{matrix}
        \hspace{0.7cm} 1 \hspace{2.3cm} \text{for $R < R_{\rm c}$} \hspace{0.5cm}  \\
        \exp \left (  - \dfrac{R-R_{\rm c}}{R_{\rm{d}}} \right ) \hspace{0.5cm} \text{for $R\geq R_{\rm c}$}
        \end{matrix}\right. \, .
        \label{HI_profile_ecuacion}
    \end{equation}
    Truncated H\,{\sc i} profiles are characteristic features of both disk and dwarf galaxies (\citealp{Swaters2002} and \citealp{, 10.1093/mnras/stw3285}). However, TDGs are typically embedded within extended H\,{\sc i} reservoirs, resulting in more uniform gas distributions and consequently flatter H\,{\sc i} profiles. For our H$\alpha$ analysis, the surface brightness profile ($\mu_{\rm{H}\alpha}$) was derived directly from optical image data. $\mu_{\rm{H}\,\text{\sc{i}}}$ and $\mu_{\rm{H}\alpha}$ are important because the effect of the interferometric beam (H\,{\sc i}) or PSF (H$\alpha$) on the measured rotation curve depends on these profiles.

\subsection{Determination of uncertainties}
\label{Determination of uncertainties}

    The parameters that explicitly influence the estimation of the dynamical mass are the radius of the disc $R_{\rm{out}}$, $R_{\rm{d}}$, $v_{\rm{rot}}$, and $\sigma_{v}$. In the case of the parametric rotation curve fit, the error of $v_{\rm{rot}}$ is given by the uncertainty in $\Sigma_{0}$ and $R_{\rm{d}}$. The uncertainties in $v_{\rm{c}}$, $M_{\rm{dyn}}$, and the ratio between the dynamic mass and baryonic mass were calculated using a Monte Carlo method. Previously, the uncertainties in $\Sigma_{0}$ and $R_{\rm{d}}$ were conservatively estimated with the differences between the values derived from the 3D fit or the rotation curve fit and those obtained from the H\,{\sc i} or optical profiles. For the error in $\sigma_{v}$, half the channel width of the H\,{\sc i} cube (6.8 km/s) was adopted, and for the fit to MEGARA data, half the spectral resolution (6.4 km/s). When $v_{\rm{rot}}$ was obtained by dividing the disk into rings, the error for each ring was obtained by calculating the standard deviation of the residuals between the velocity maps ($M_{1}$ in Eq. \ref{ecuaciones_momentos}) of the data and the fit, which is then divided by $\sin i$ (consistently with Eq. \ref{vrot2D}). The error in $R_{\rm out}$ is half the width of the last ring.

    The uncertainties in $v_{\rm rot}$ presented in Figs. \ref{rotation_dispersion_arp72b_new_version} and \ref{rotation_dispersion_arp72c_new_version} were derived from the $v_{\rm obs}$ errors using Eq. \ref{vrot2D}, where the latter were estimated by applying a Monte Carlo method at each position.

\begin{figure*}[ht]
    \centering

    \begin{subfigure}[b]{0.9\linewidth}
        \centering

        \includegraphics[width=\linewidth]{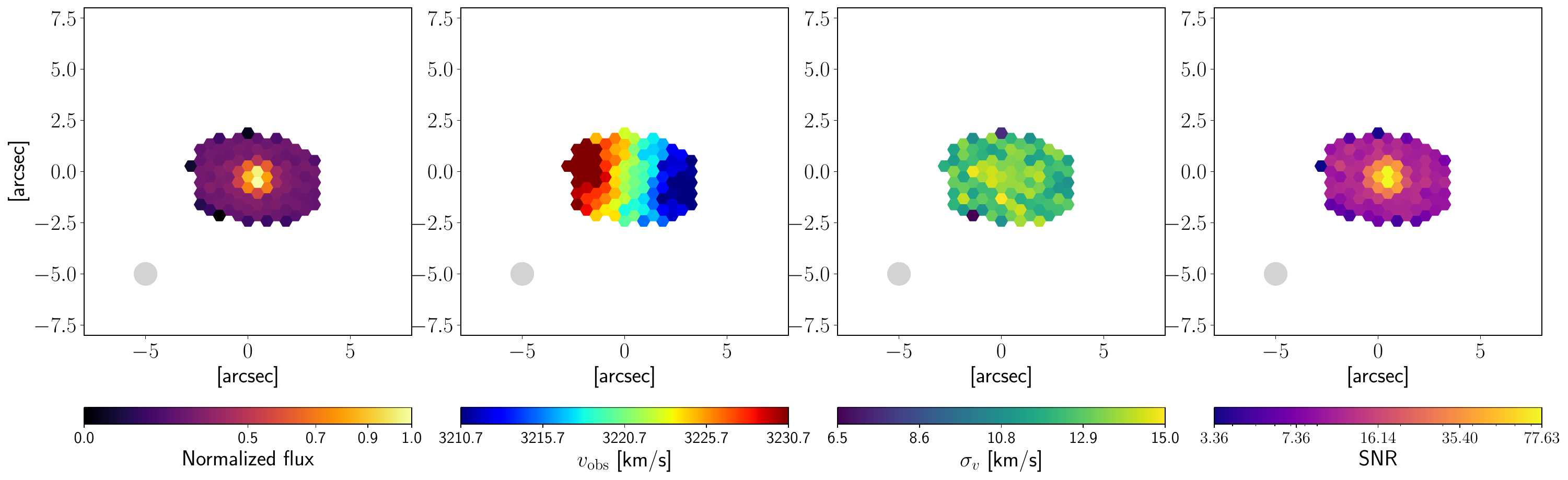}

    \end{subfigure}
    \hfill 
    \begin{subfigure}[b]{0.9\linewidth}
        \centering

        \includegraphics[width=\linewidth]{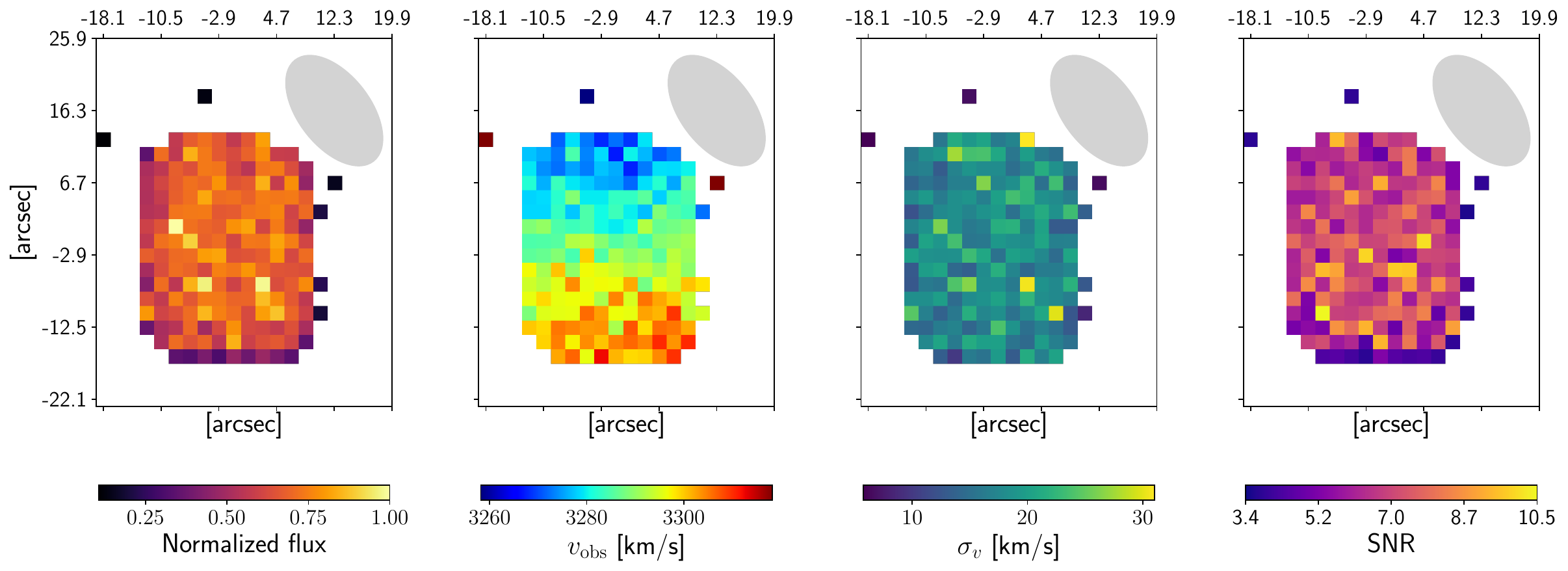}

    \end{subfigure}

    \caption{Examples of synthetic observations used to determine the uncertainties of the fitted parameters. Top panels: Representative simulation of fiber-based data with characteristics matching the MEGARA observations. Bottom panels: Representative simulation of a data cube resembling the GMRT observations. From left to right, the panels display the intensity, velocity, velocity dispersion, and signal-to-noise ratio (SNR) maps. The gray ellipse and gray circle in each panel indicates the GMRT synthesized beam or the MEGARA PSF, as appropriate.}
    \label{GMRT_MEGARA_example_simulations}
    \end{figure*}

\begin{figure*}[hb]
    \centering
    
    \begin{subfigure}[b]{0.48\linewidth}
        \centering
        
        \includegraphics[width=\linewidth]{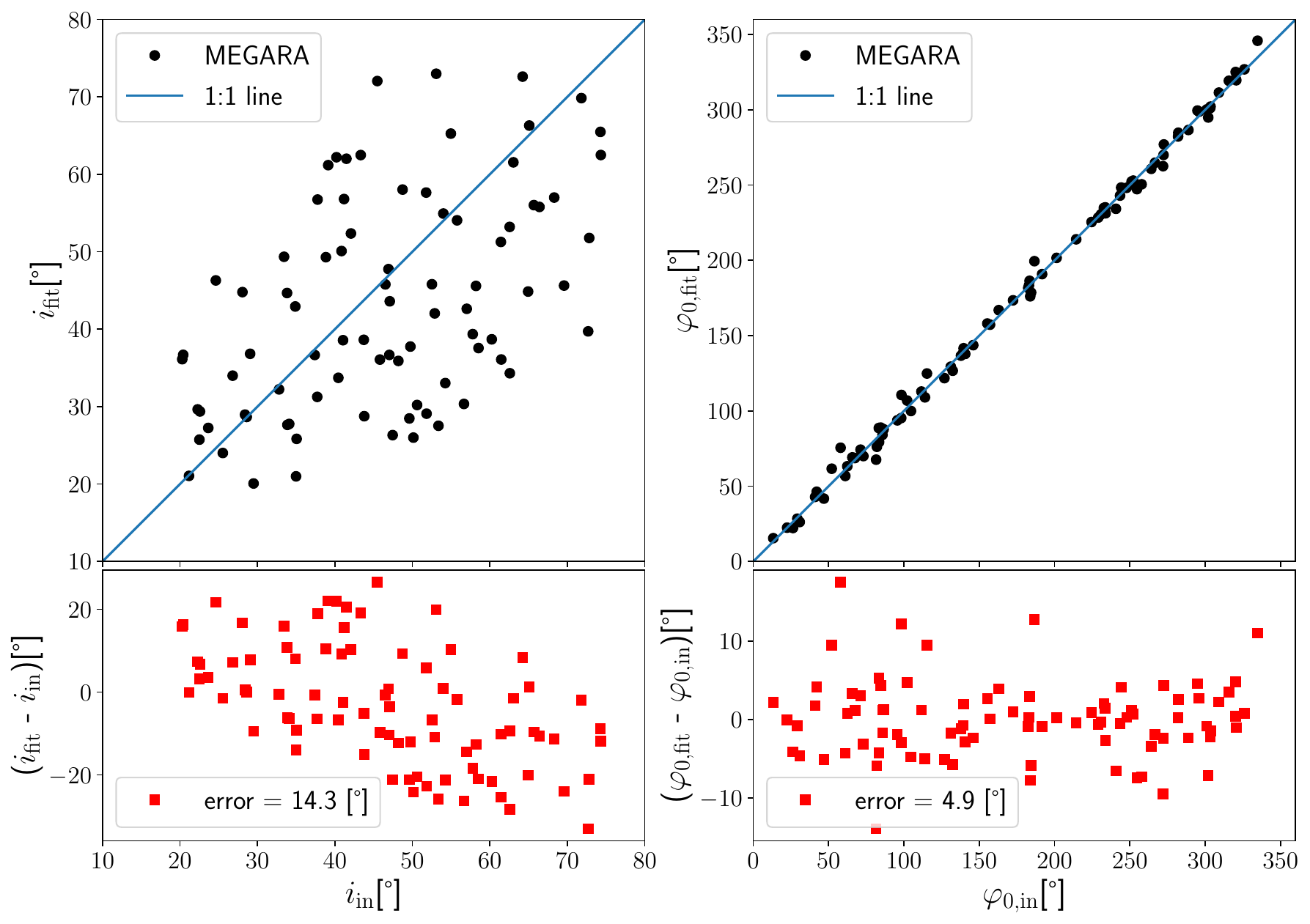}
        
    \end{subfigure}
    \hfill 
    \begin{subfigure}[b]{0.48\linewidth}
        \centering
        
        \includegraphics[width=\linewidth]{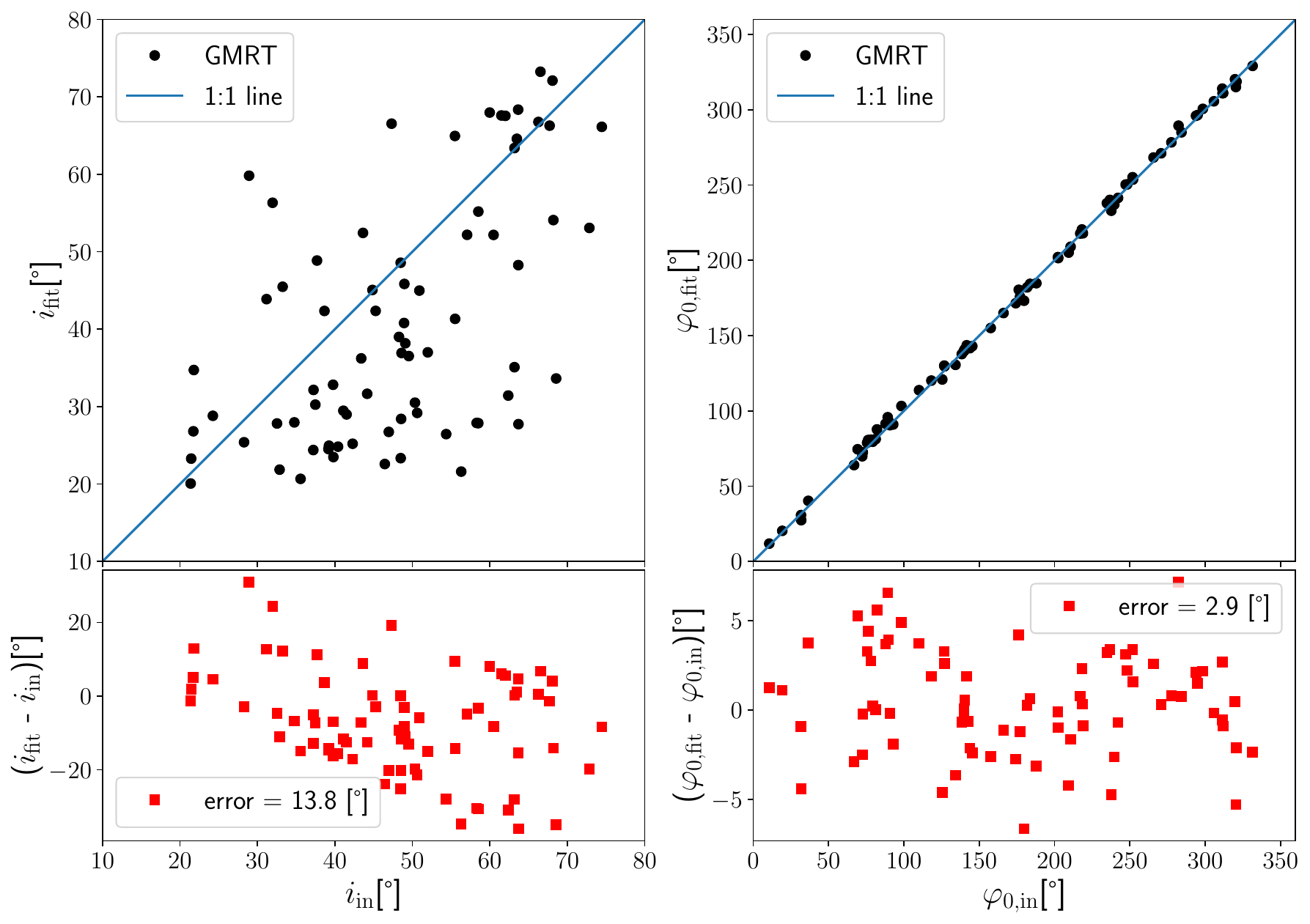}
        
    \end{subfigure}
    
    \caption{Fitted parameters versus input model parameters for $i$ and $\varphi_{0}$. The left panels correspond to data for the MEGARA observations, while the right panels represent data for the GMRT observations. For each case, the lower subpanels show the residuals (the difference between the fitted and input parameters), and the uncertainty for each parameter was calculated as the standard deviation of these residuals.}
    \label{simulations GMRT y MEGARA}
    \end{figure*}

    $\varphi_{0}$ and $i$ are two fundamental parameters in the kinematic modeling of rotating galaxies, as they define the observed velocity field. Among these, $i$ is particularly challenging to constrain due to its intrinsic degeneracy with the rotational velocity. To estimate the uncertainties for these parameters, we generated synthetic observations using parameter combinations drawn from random distributions. Noise was added to these models to replicate the SNR maps of the actual observations. These synthetic cubes were then processed (line detection at each position) and fitted following the same methodology applied to the real data. Representative examples of these simulations, including moment and SNR maps, are presented in Fig. \ref{GMRT_MEGARA_example_simulations}. For the input rotation velocities, the distributions were constrained to values typical of TDGs. The total number of simulations was limited by computational cost. To ensure a robust error estimation, we discarded fits where the residual ($|{\rm par}_{\rm in} - {\rm par}_{\rm fit}|$) was an outlier exceeding $2\sigma_{\rm par}$, where $\sigma_{\rm par}$ represents the standard deviation of the parameter residuals. The resulting error distributions for $\varphi_{0}$ and $i$ are shown in Fig. \ref{simulations GMRT y MEGARA} right for data cubes like GMRT observations and in Fig. \ref{simulations GMRT y MEGARA} left for data organized into fibers, as in the case of MEGARA. The final uncertainties for $\varphi_{0}$, $i$, and $v_{\rm sys}$ reported in the parameter tables were derived from this methodology.

    In this work, the uncertainty in $v_{\rm rot}$ is reported based on the residuals between the observed velocity maps and the best-fit models, which characterizes the extent to which the object deviates from an ideal rotating disk. Other authors, such as \cite{Lelli2015}, estimate the error in $v_{\rm rot}$ using Eq. \ref{lellierror}; this approach accounts for the influence of the inclination uncertainty, since $v_{\rm rot} \propto v_{\rm l.o.s} / \sin i$, where $v_{\rm l.o.s}$ is the projected line-of-sight velocity. Following their methodology, we adopted $\sigma_{v_{\rm l.o.s}}$ as half the channel width for H\,{\sc i} and half the spectral resolution for H$\alpha$, combined with the $\sigma_{i}$ values derived in our previous analysis. We then performed a Monte Carlo error propagation to evaluate the impact of $\sigma_{i}$ on the $M_{\rm dyn} / M_{\rm bary}$ ratio. These results are summarized in Table \ref{Tabla_resultados_general_lelli_error}.

    \begin{equation}
        \Delta v_{\rm error} = \sqrt{\left [ \frac{\Delta v_{\rm l.o.s}}{\sin (i)} \right]^{2}+\left [ v_{\rm rot}\frac{\Delta i_{\rm error}}{\tan (i)} \right]^{2}}
        \label{lellierror}
    \end{equation}

    \begin{table}[!htb]
    \caption{Uncertainties based on Eq. \ref{lellierror}.}
    \centering
    \begin{adjustbox}{width=\linewidth,center}

    \begin{tabular}{ccccc}
    
    \hline
    \hline
    
    Object & $\Delta v_{\rm rot-error}$ [km/s] & $v_{\rm c}$ [km/s] & $M_{\rm{dyn}}$ {[}$10^{8}$M$_{\odot}${]} & $\dfrac{M_{\rm{dyn}}}{M_{\rm{bary}}}$ \\ \hline \hline

    Arp\,72b - H${\alpha}$ & 14 & $16^{+16}_{-3}$ & $0.4^{+1.1}_{-0.1}$ & $0.7^{+2.2}_{-0.2}$ \\ \hline

    Arp\,72c - H\,{\sc i} & 9.4 & $21^{+10}_{-6}$ & $4.6^{+5.3}_{-2.2}$ & $1.1^{+1.2}_{-0.5}$ \\ \hline

    \end{tabular}
    \end{adjustbox}
    \tablefoot{Uncertainty in the rotational velocity derived using Eq. \ref{lellierror} and its impact on the error estimates for $v_{\rm c}$, $M_{\rm dyn}$, and the $M_{\rm dyn}/M_{\rm bary}$ ratio.}
    \label{Tabla_resultados_general_lelli_error}
    \end{table}

    The results presented in Table \ref{Tabla_resultados_general_lelli_error} indicate that, only for Arp\,72b, the upper uncertainty increases significantly compared to the values reported in Table \ref{Tabla_resultados_general} for $v_{\rm c}$, $M_{\rm dyn}$, and $M_{\rm dyn}/M_{\rm bary}$. This increased uncertainty is driven by the specific inclination angle of Arp\,72b. Across all results in this study, the $M_{\rm dyn}/M_{\rm bary}$ ratio consistently exhibits a larger positive uncertainty, such that the upper-limit values for $M_{\rm dyn}/M_{\rm bary}$ are approximately 2. Any apparent excess in dynamical mass could potentially be accounted for by the current lack of molecular gas measurements.
    
    Furthermore, the synthetic observation fits allowed us to estimate the uncertainties for the remaining model parameters, which are reported in Table \ref{Tabla_incertidumbres_models}. For the MEGARA data (fixed-ring method), $\Delta v_{\rm rot}$ was calculated based on the residuals of $v_{\rm rot}$ across all fitted rings.

    \begin{table}[!htb]
    \caption{Results of fits to synthetic data.}
    \centering
    \begin{adjustbox}{width=\linewidth,center}
    \begin{tabular}{ccccc}
    
    \hline
    \hline

    Parameter & $\Delta \sigma_{v}$ [km/s] & $\Delta R_{\rm d}$ [kpc] & $\Delta \Sigma_{0}$ {[}M$_{\odot}$/pc$^{2}${]} & $\Delta R_{\rm c}$ [kpc] \\ \hline \hline

    error$_{\rm GMRT}$ & 1 & $2.7$ & $14$ & 1 \\ \hline \hline
    Parameter & $\Delta \sigma_{v}$ [km/s] & $\Delta v_{\rm rot}$ [km/s]  \\ \hline

    error$_{\rm MEGARA}$ & 1 & 6.8 &   \\ \hline
    
    \end{tabular}
    \end{adjustbox}
    \tablefoot{Uncertainties for the remaining parameters derived from fits to synthetic observations.}
    \label{Tabla_incertidumbres_models}
    \end{table}
    
\FloatBarrier
\section{3D fitting visualization}

    The 3D fitting results are presented through two complementary visualization methods: moment maps (showing intensity, velocity, and velocity dispersion) and PV (Position-Velocity) diagrams along the kinematic axes.
    
    \begin{figure*}[hb]
    \centering

    \begin{subfigure}[b]{0.7\linewidth}
        \centering

        \includegraphics[width=0.9\linewidth]{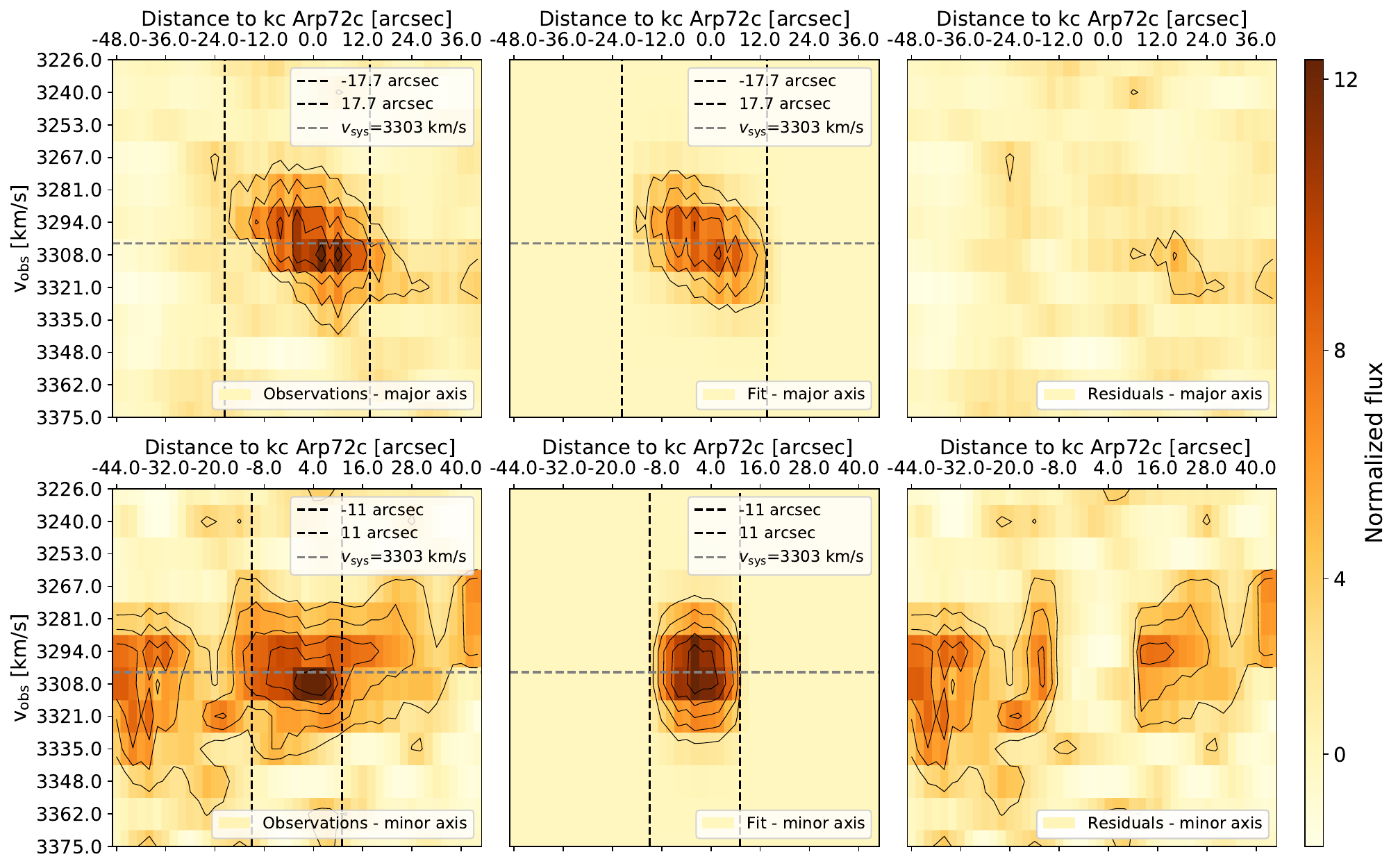}

    \end{subfigure}
    \hfill 
    \begin{subfigure}[b]{0.29\linewidth}
        \centering

        \includegraphics[width=\linewidth]{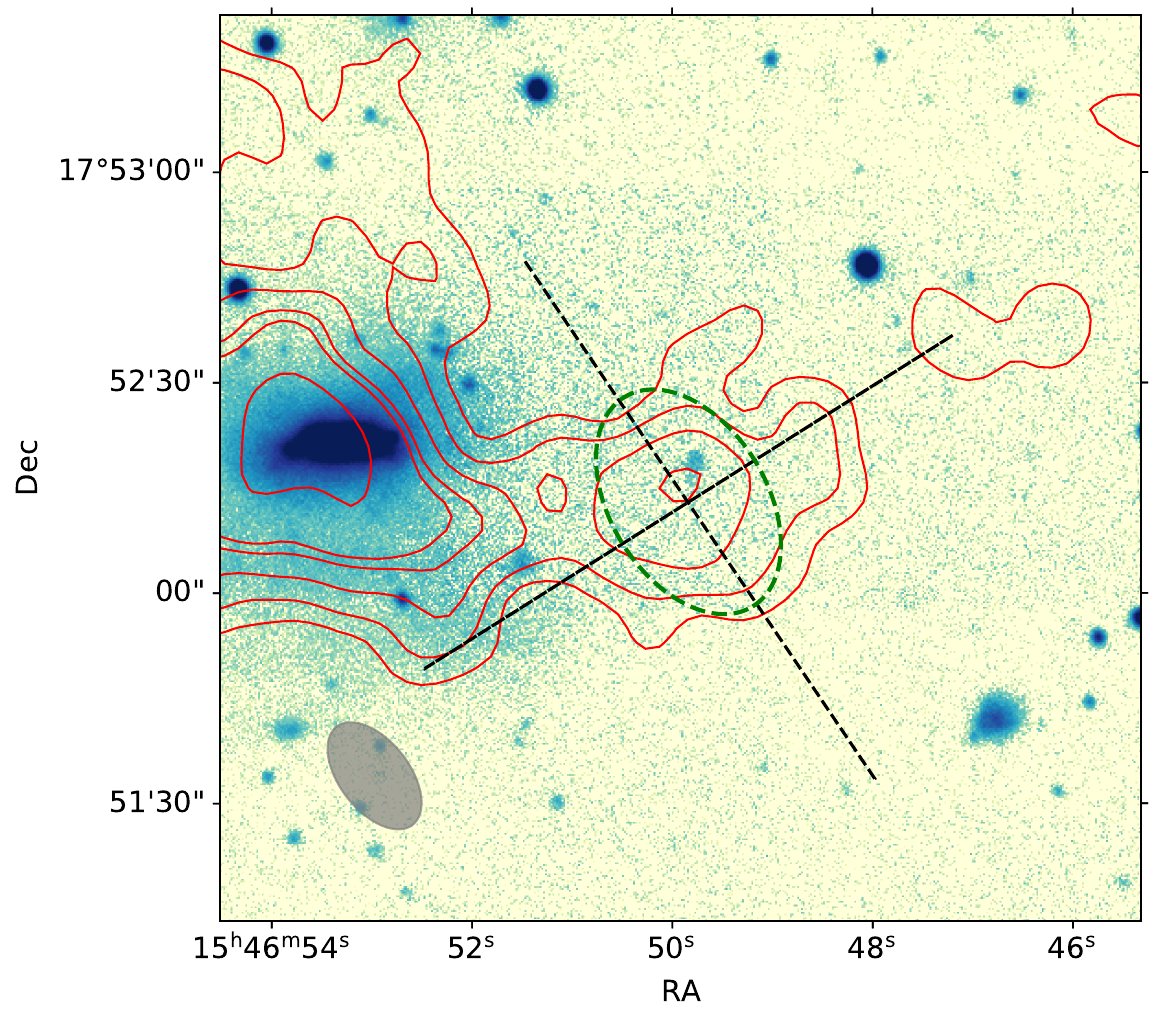}

    \end{subfigure}
    
    \caption{PV (Position-Velocity) diagram of Arp\,72c - H\,{\sc i} constructed using the kinematic axes given by the position angle ($\varphi_{0}$) and the kinematic center of the 3D fit shown in the figure on the right with dashed black lines. The image shows the DESI $g$ filter, along with HI contours in red and the 3D fit disk within the dashed green ellipse. Left figure, first column: PV diagrams obtained from the spectral cube of the GMRT data. Second column: PV diagrams obtained from the spectral cube of the best fit. Third column: residuals for each kinematic axis. The PV diagrams were normalized with respect to the amplitude of the line at the position of the kinematic center, and the contours are $(3, 5, 7, 9, 11)\times 0.0057$\,Jy/beam.}
    \label{fig:Arp72cHI_analysis}
    \end{figure*}
    
    \begin{figure*}[!ht]
        \centering
        \includegraphics[width=0.98\hsize]{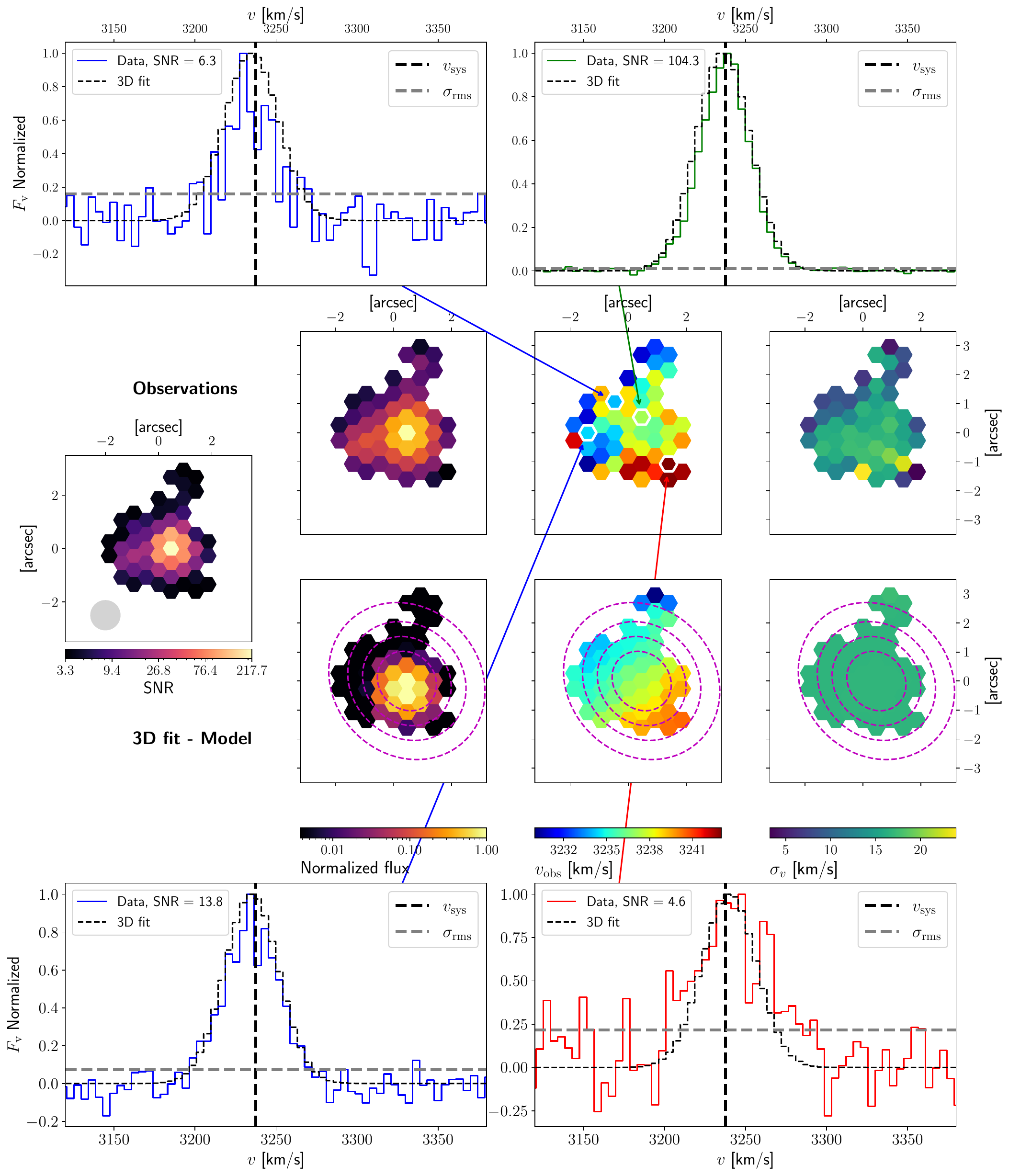}
        \caption{Moment maps and 3D fit results to the H$\alpha$ data for Arp\,72b. Middle panels: Intensity, velocity, and velocity dispersion maps for the MEGARA IFU data (top) and best-fit model (bottom). The SNR map and observational PSF are shown on the left. Model panels: Dashed magenta circles indicate the rings used in the fixed-ring method. Intensity is normalized to the flux at the kinematic center. Top and bottom panels: Comparison between observed and modeled spectra at the marked positions. Dashed black vertical and gray horizontal lines represent the systemic velocity and the noise level, respectively.}
        \label{3Dfit_arp72b_new_version}
    \end{figure*}

    \begin{figure*}[!ht]
        \centering
        \includegraphics[width=0.98\hsize]{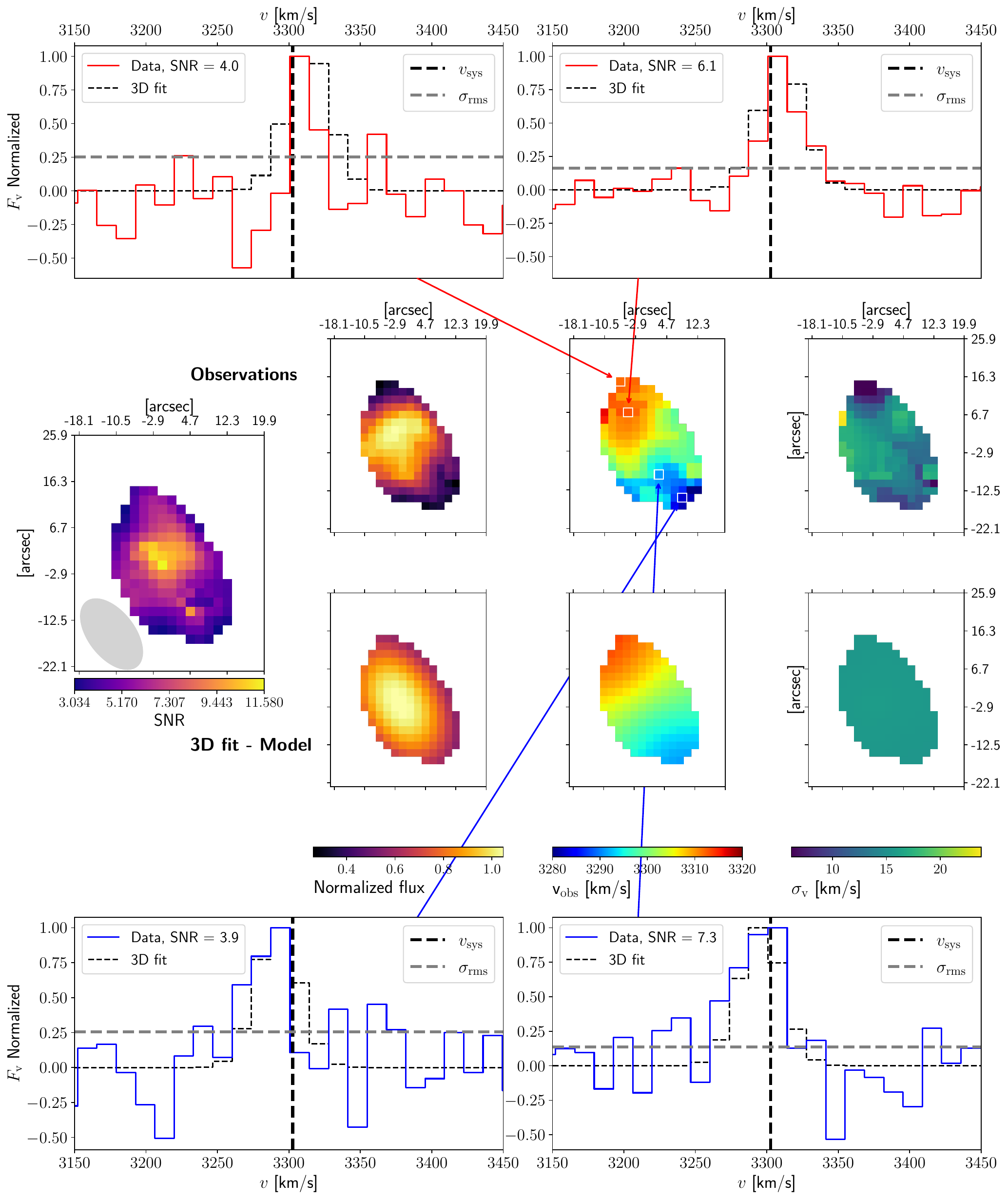}
        \caption{Moment maps and 3D fit results to the H\,{\sc i} data for Arp\,72c. Middle panels: Intensity, velocity, and velocity dispersion maps for the GMRT data (top) and best-fit model (bottom). The SNR map and size beam are shown on the left. Intensity map is normalized to the flux at the kinematic center. Top and bottom panels: Comparison between observed and modeled spectra at the marked positions. Dashed black vertical and gray horizontal lines represent the systemic velocity and the noise level, respectively.}
        \label{3Dfit_arp72c_new_version}
    \end{figure*}

\subsection{Arp\texorpdfstring{\,}{ }72b - {\rm H}\texorpdfstring{$\alpha$}{alfa} fit}
\label{moments fit Arp72b}
    The intensity map in Fig. \ref{3Dfit_arp72b_new_version} shows extremely compact emission. Even when using only the inner exponential profile from Fig. \ref{F606W_profiles} in our disk model, the H$\alpha$ appears slightly more extended.
    
    In Fig. \ref{3Dfit_arp72b_new_version} the variations in the dispersion map of our fit are very small (dominated by beam smearing) compared to the observations. This discrepancy arises because our fit assumes a uniform $\sigma_{v}$ across the disk, whereas the $\sigma_{v}$ maps exhibit notable deviations. On the other hand, we found $\sigma_{v}$$\sim$16.6 km/s in the fit and $\sigma_{v, \, {\rm 3D-fit}} = 14.4$ km/s for the disk, showing that the effects of LSF and PSF give an increase of 2.2 km/s in the observations of velocity dispersion.

\subsection{Arp\texorpdfstring{\,}{ }72c - \texorpdfstring{{\rm H}\,{\sc i}}{HI} fit}
\label{moments and pv Arp72c}
    
    The velocity map within the red ellipse in Fig. \ref{all_optical_imaages}, right panel, displays clear rotational signatures (Fig. \ref{3Dfit_arp72c_new_version}, observations velocity map). We performed the 3D fit by directly parameterizing the rotation curve using $\Sigma_{0}$ and $R_{\rm{d}}$, though this approach systematically overestimates the velocity dispersion in regions distant from the kinematic center. While the model yields relatively uniform velocity dispersion values ($\sigma_{v}$$\sim$15.7 km/s) across the disk, the observed data show significantly greater variation. The quality of the fit is more easily assessed using the PV diagrams in Fig. \ref{fig:Arp72cHI_analysis}, left panel. Along the major axis, our model produces minimal residuals, indicating a good fit. However, along the minor axis, the PV diagram follows the tidal tail structure, resulting in significant H\,{\sc i} residuals on both sides of the kinematic center that clearly lie outside the disk component, though the central regions show good agreement. The clearest differences between the model and the observations are in the intensity map, where in the observations the distribution of H\,{\sc i} is oriented approximately along the minor axis of the disk. This result is natural because the tidal tail goes in that direction, so there will be more material in this direction (Fig. \ref{all_optical_imaages}, right panel showing H\,{\sc i} contours overlaid on the optical counterpart).

\FloatBarrier 
\clearpage

\end{appendix}

\end{document}